\documentclass{aa}  
\usepackage{graphicx}
\usepackage{natbib}
\usepackage{amssymb}
\usepackage{amsmath}
\usepackage{amstext}
\usepackage{rotate}
\usepackage{rotating}
\usepackage{supertabular}
\usepackage{epsfig}
\usepackage{lscape}
\usepackage{enumerate}

\bibpunct{(}{)}{;}{a}{}{,}
\usepackage{txfonts}
\newcommand{\gsim}{\hbox{\rlap{\lower.55ex\hbox{$\sim$}} \kern-.3em
\raise.4ex \hbox{$>$}}}
\newcommand{\lsim}{\hbox{\rlap{\lower.55ex\hbox{$\sim$}} \kern-.3em
\raise.4ex \hbox{$<$}}}

\newcommand{\nha}{\textsc{[N\,ii]}$\lambda$6584/H$\alpha$}
\newcommand{\sha}{\textsc{[S\,ii]}$\lambda\lambda$6717,6731/H$\alpha$}

\newcommand{\ohb}{\textsc{[O\,iii]}$\lambda$5007/H$\beta$}
\newcommand{\hb}{H$\beta$}
\newcommand{\hg}{H$\gamma$}
\newcommand{\ha}{H$\alpha$}

\newcommand{\hi}{\textsc{H\,i}}
\newcommand{\hei}{He\,\textsc{i}}
\newcommand{\siii}{[S\,\textsc{iii}]}
\newcommand{\oii}{[O\,\textsc{ii}]}
\newcommand{\neiii}{[Ne\,\textsc{iii}]}
\newcommand{\sii}{\textsc{[S\,ii]}}
\newcommand{\oiiB}{O\,\textsc{ii}}
\newcommand{\feii}{[Fe\,\textsc{ii}]}
\newcommand{\feiii}{[Fe\,\textsc{iii}]}
\newcommand{\cii}{C\,\textsc{ii}}
\newcommand{\oi}{O\,\textsc{i}}
\newcommand{\mgi}{Mg\,\textsc{i}]}
\newcommand{\ariv}{[Ar\,\textsc{iv}]}
\newcommand{\siiiB}{\textsc{Si\,ii}}
\newcommand{\nii}{\textsc{[N\,ii]}}
\newcommand{\ariii}{[Ar\,\textsc{\,iii]}}
\newcommand{\oiii}{\textsc{[O\,iii]}}

\newcommand{\oiiioii}{\textsc{[O\,iii]}$\lambda\lambda$4959,5007/\textsc{[O\,ii]}$\lambda\lambda$3726,3729}
\newcommand{\ghiir}{GH\,\textsc{ii}R}

\begin{document}
%
   \title{The ionized gas in the central region of \object{NGC~5253}:}
\subtitle{2D mapping of the physical and chemical properties\thanks{Based on observations  
       collected at the European Organisation for Astronomical
       Research in the Southern Hemisphere, Chile (ESO Programme
       078.B-0043 and 383.B-0043).}} 
   \author{A. Monreal-Ibero\inst{1}
          \and
         J. R. Walsh\inst{2}
         \and
         J. M. V\'{\i}lchez\inst{1} 
         }

   \offprints{A. Monreal-Ibero}

   \institute{Instituto de Astrof\'{\i}sica de Andaluc\'{\i}a (CSIC), C/
              Camino Bajo de Hu\'etor, 50, 18008 Granada, Spain. 
              \email{ami@iaa.es}
   \and
   European Southern Observatory, Karl-Schwarzschild Strasse 2, D-85748 Garching bei M\"unchen, Germany.
} 
   \date{Received: 4 May 2012 / Accepted: 8 June 2012 }

 
  \abstract
   {Blue Compact Dwarf (BCD) galaxies constitute the ideal laboratories to test the interplay between massive star formation and the surrounding gas. As one of the nearest BCD galaxies, NGC~5253 was previously studied with the aim to elucidate in detail the starburst interaction processes. Some open issues regarding the properties of its ionized gas still remain to be addressed.} 
   {The 2D structure of the main physical and chemical properties of the ionized gas in the core of NGC~5253 has been studied.}  
   {Optical integral field spectroscopy (IFS) data has been obtained with FLAMES Argus and lower resolution gratings of the Giraffe spectrograph.}
   {
We derived 2D maps for different tracers of electron density ($n_e$), electron temperature ($T_e$) and 
ionization degree. The maps for $n_e$ as traced by \oii, \sii, \feiii, and \ariv\, line ratios are compatible 
with a 3D stratified view of the nebula with the highest $n_e$ in the innermost layers and a decrease of 
$n_e$ outwards.
To our knowledge, this is the first time that a $T_e$ map based on \sii\ lines for an extragalactic object is presented.
The joint interpretation of the $T_e$(\sii) and $T_e$(\oiii) maps is consistent with a $T_e$ structure in 3D 
with higher temperatures close to the main ionizing source surrounded by a colder and more diffuse component.  
The highest ionization degree is found at the peak of emission for the gas with relatively high ionization in 
the main \ghiir\ and lower ionization degree delineating the more extended diffuse component. 
We derived abundances of oxygen, neon, argon, and nitrogen.
Abundances for $O$, $Ne$ and $Ar$ are constant over the mapped area within $\lsim$0.1~dex. The mean $12+\log(O/H)$ is $8.26\pm0.04$ while the 
relative abundances of $\log(N/O)$, $\log(Ne/O)$ and $\log(Ar/O)$ were $\sim-1.32\pm0.05$, $-0.65\pm0.03$ 
and $-2.33\pm0.06$, respectively.
There are two locations with enhanced $N/O$. The first ($\log(N/O)\sim-0.95$) 
occupies an area of about 80~pc$\times$35~pc and is associated to two super star clusters. The second 
($\log(N/O)\sim-1.17$), reported here for the first time, is associated to two moderately massive 
($2-4\times10^4$~M$_\odot$) and relatively old ($\sim10$~Myr) clusters.
A comparison of the $N/O$ map 
with those produced by strong line methods supports the use of N2O2 over N2S2 in the search for chemical 
inhomogeneities \emph{within} a galaxy. The results on the localized nitrogen enhancement were used to 
compile and discuss the factors that affect the complex relationship between Wolf-Rayet stars and $N/O$ excess.
%
%
%
%
}  
 {}


   \titlerunning{Physical and chemical conditions of the ionized gas in NGC~5253}
   \maketitle
%

\section{Introduction}

Local Blue Compact Dwarf galaxies \citep[BCDs, see ][for a review]{kun00} constitute ideal laboratories to test how the interaction between massive star formation, gas and dust affects galaxy evolution. Specifically, both winds of massive stars and supernova explosions i) inject mechanical energy to the interstellar medium (ISM), redistributing the gas within the galaxy and therefore, quenching (or igniting) future star formation; ii) eject processed material into the ISM thus causing a chemical enrichment of the galaxy.

NGC~5253 is an example BCD particularly suited for the study of the interaction between gas and massive star formation, since it is relatively close and a wealth of ancillary information is available in all spectral ranges, from X-ray to radio.
This galaxy belongs to the Centaurus~A / M~83 complex \citep{kar07} and is suffering a recent burst of star formation. The existence of an \textsc{H\,i} plume extending along the optical minor axis \citep{kob08} supports the idea of the burst being caused by a former encounter with M~83 \citep{van80}. 
A wealth of studies at different wavelengths shows that this galaxy is peculiar in several aspects. In particular, high resolution multiband photometry with the \emph{Hubble Space Telescope} (HST) revealed the existence of several relatively young star clusters in the central region of the galaxy with typical masses of $\sim2-120\times10^3$~M$_\odot$ \citep[e.g.][]{har04}. Among those, two very massive ($\sim1-2\times10^6$~ M$_\odot$) Super Star Clusters (SSCs) at the nucleus of the galaxy, separated by $\sim$0\farcs4 \citep{alo04} and associated with a very dense compact H\,\textsc{ii} region, only detected in the radio at 1.3~cm and 2~cm \citep{tur00}, stand out as candidates to be the youngest globular cluster(s) yet observed \citep{gor01}. Also, \citet{lop12} showed how this galaxy does not satisfy the Schmidt-Kennicutt law of star formation and seems to be slightly metal-deficient in comparison with starbursts of similar baryonic mass. But probably, the most well-known peculiarity of NGC~5253 is the existence of areas with enhanced abundance of nitrogen. This was first reported by \citet{wel70} and confirmed afterwards by several works \citep[e.g.][]{wal89}.

Narrow band imaging showed that as a consequence of its starburst nature, the ionized gas in \object{NGC~5253} presents a complex structure that includes filaments and arcs \citep[e.g.][]{cal04}. Therefore, proper characterization of the properties of the ionized gas in \object{NGC~5253} would benefit from high quality two-dimensional spectral mapping over a contiguous area of interest. Nowadays, Integral Field Spectroscopy (IFS) facilities represent the obvious choice to obtain this kind of information and several works devoted to the study of BCDs using this approach have been published in recent years \citep[e.g.][]{izo06,gar08,jam09,wes10}.

\begin{figure}[th]
   \centering
\includegraphics[angle=0,width=0.45\textwidth, clip=,bb = 170 50 430 340]{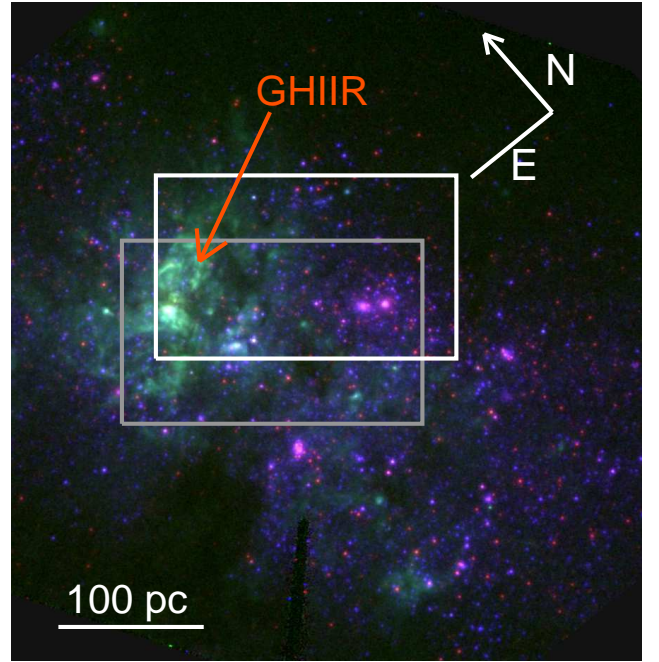}
   \caption[Covered field with FLAMES]{False colour image in filters
     F435W ($B$, blue channel), F658N ($H\alpha$, green channel), and
     F814W ($I$, red channel) for \object{NGC~5253} using HST-ACS images
     (programme 10608, P.I.:Vacca). The area covered in ESO program
     383.B-0043 (078.B-0043) is marked with a white (grey) rectangle     while the main \ghiir\ of the galaxy is labeled in orange. 
     The orientation and
     scale for a distance of 3.8~Mpc are indicated. \label{apuntado}} 
 \end{figure}

\defcitealias{mon10}{Paper~I}
In particular, \citet[][hereafter Paper I]{mon10} carried out a detailed study of the central region in \object{NGC~5253} using IFS data collected with FLAMES \citep{pas02}.
One of the topics studied in more detail there was the kinematics of the ionized gas. We found that the line profiles were complex, needing up to three components to reproduce them. In the main Giant H\,\textsc{ii} Region (\ghiir, see Fig. \ref{apuntado}), one of them presented supersonic widths while the other two were relatively narrower. Moreover, the broad component presented and excess in nitrogen of $\sim$1.4 times than that of the narrow one \citep{mon11b}. This was consistent with a scenario where the two SSCs produce an outflow that encounters the previously quiescent gas.
Also, we delimited very precisely the area polluted with extra nitrogen, based on the excess of the \nha\ ratio with respect to \sha\ and clearly demonstrated that, at least, some of the Wolf-Rayet (WR) star population cannot be the cause of this enrichment. Moreover, we could resolve a long-standing issue regarding the elusive He\,\textsc{ii} emission at $\lambda$4686~\AA. \citet{cam86} mentioned a possible detection of He\,\textsc{ii} but this result was not certainly confirmed afterwards. We detected several localized areas with clear He\,\textsc{ii}$\lambda$4686 detection although, rather puzzlingly, not coincident in general with the area exhibiting extra nitrogen.

With the limited spectral range utilized in that work, we could only determine \emph{relative} nitrogen abundances. Moreover, while interpreting the increase of \nii$\lambda$6584/\sii$\lambda\lambda$6717,6731 as caused by an enhancement in nitrogen abundance is the most natural explanation and is supported by previous observational evidence \citep[e.g.][]{kob97}, this is not the only alternative. Instead, specific combinations of ionization structure and metallicity gradient could also produce similar observables. Since dwarf galaxies are not expected to have strong metallicity gradients in a similar manner to large spirals, this option seems unlikely. However, it could not be rejected \emph{a priori} in our previous work. 

Here, we present the natural continuation of the work in \citetalias{mon10}. To overcome the mentioned drawbacks, we obtained new FLAMES observations that allow us to map missing physical properties such as electron temperature ($T_e$) and local degree of excitation. In this manner we are able to: i) map the chemical content in the central part of the galaxy and detect inhomogeneities (if any) in an unbiased manner; ii) evaluate how well relative abundance tracers based on strong emission lines reproduce the values derived from direct measurements; iii) check if the derived physical and chemical structure is consistent with the picture sketched in \citetalias{mon10}.

The paper is organized as follows: Sec. \ref{obsred} describe the observations and technical details about the data reduction and processing necessary to extract the required information for the analysis. Sec. \ref{resultados} contains the results from one Gaussian fitting for the physical (i.e. $T_e$, $n_e$, ionization degree) and chemical (i.e. metallicity and relative abundances) properties both in 2D and in particularly interesting apertures. Sec. \ref{discusion} will explore how the derived maps trace the 3D physical structure of the central part of \object{NGC~5253}, and the reliability (or limitations) of strong line based tracers to determine the relative abundance in nitrogen.
Basic information for \object{NGC~5253} can be found in Table~1 in \citetalias{mon10}.


\section{Observations and data processing \label{obsred}}

\subsection{Observations}

Data were obtained with the \emph{Fibre Large Array Multi Element
  Spectrograph}, FLAMES \citep{pas02} at Kueyen, Telescope Unit 2
  of the 8~m VLT at ESO's observatory on Paranal. We used the ARGUS
  Integral Field Unit with the sampling of
  0.52$^{\prime\prime}$/lens. This permits coverage of a field of view (fov) of
  $11\farcs5 \times 7\farcs3$. Utilized gratings were L385.7 (LR1), L427.2 (LR2), and L479.7 (LR3) and L682.2 (LR6).

Data for the two last gratings were obtained in visitor mode on February 10, 2007 (programme 078.B-0043).
Details of these observations appear in \citetalias{mon10}. In addition to these, data for the L385.7 (LR1) and  L427.2 (LR2) were obtained in service mode during June 2009 (programme 383.B-0043). The spectral range, resolving power, exposure
  time and airmass for each configuration are listed in Table
  \ref{log_observaciones}. Seeing ranged typically between $0\farcs4$
  and $1\farcs2$ with a median($\pm$standard deviation) of $\sim0\farcs8$ $(\pm0\farcs2)$.  
  All the data were taken under clear conditions. In addition,
  standard sets of calibration files were obtained. These included
  continuum and ThAr arc lamps exposures as well as 
  frames for the spectrophotometric standard stars HR~7596 and
  LTT~3218 for the L385.7 (LR1) and L427.2 (LR2) gratings respectively.

Due to a guiding problem, there was a $\sim2\farcs9$ offset between
the pointings on runs 078.B-0043 and 383.B-0043. In spite of it, the main emitting
regions were covered by all the configurations. The
precise area covered in each run is shown in Figure \ref{apuntado}
which contains the FLAMES fov over-plotted on an HST $B$, \ha,
$I$ colour image.

\begin{table}
\centering
      \caption[]{Observation log \label{log_observaciones}}
              \begin{tabular}{ccccccccc}
            \hline
            \noalign{\smallskip}
Grating & Spectral range & Resolution & t$_{\mathrm{exp}}$ & Airmass\\
        &  (\AA)         &              & (s)        &      \\
            \noalign{\smallskip}
            \hline
            \noalign{\smallskip}
L385.7    & 3\,610--4\,081 & 12\,800 & $21\times895$ &  1.00--1.10\\
L427.2    & 3\,964--4\,567 & 10\,200 & $9\times895$  &  1.00--1.09\\
            \noalign{\smallskip}
            \hline
         \end{tabular}
\end{table}

   \begin{figure*}[t!]
   \centering
\includegraphics[angle=0,width=0.45\textwidth, clip=,bb = 110 360 900 870]{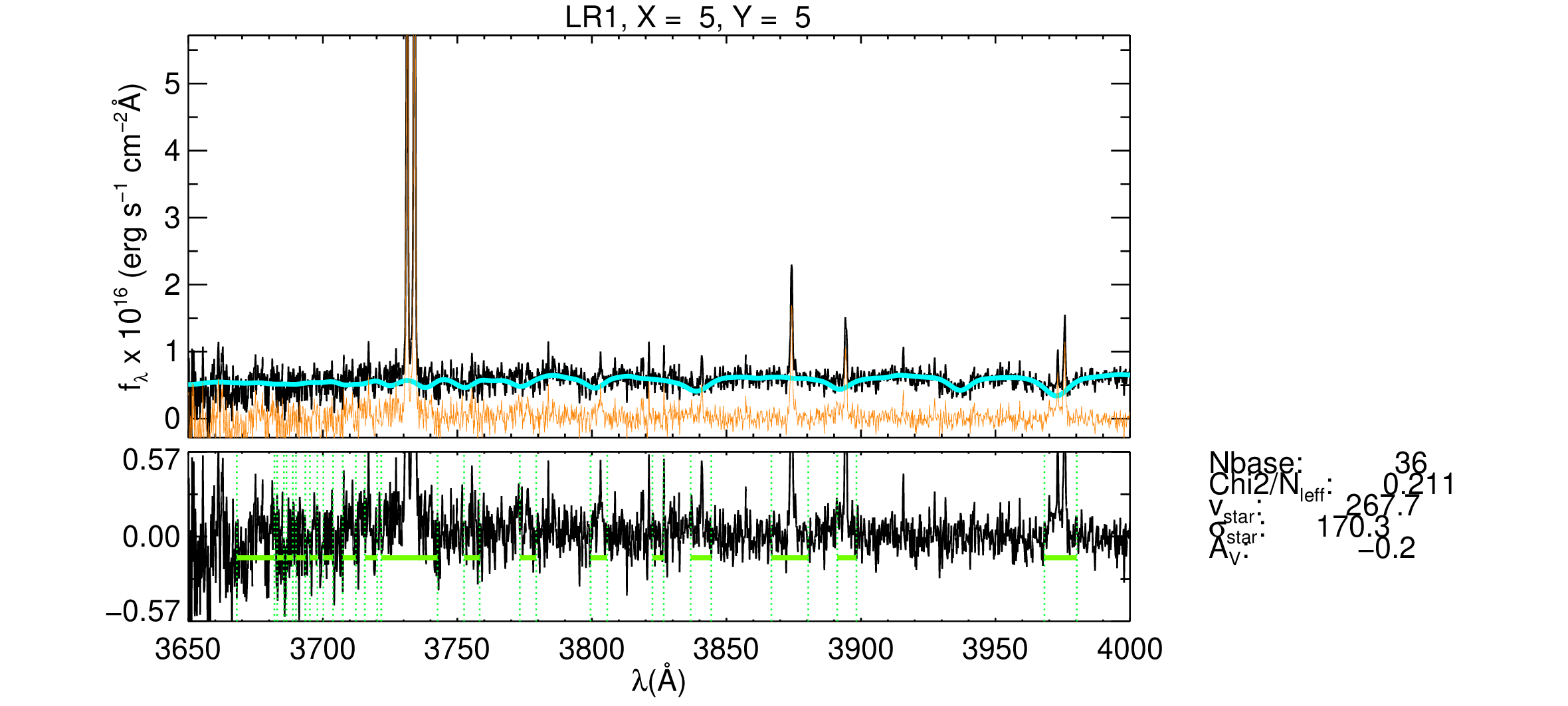}
\includegraphics[angle=0,width=0.45\textwidth, clip=,bb = 110 360 900 870]{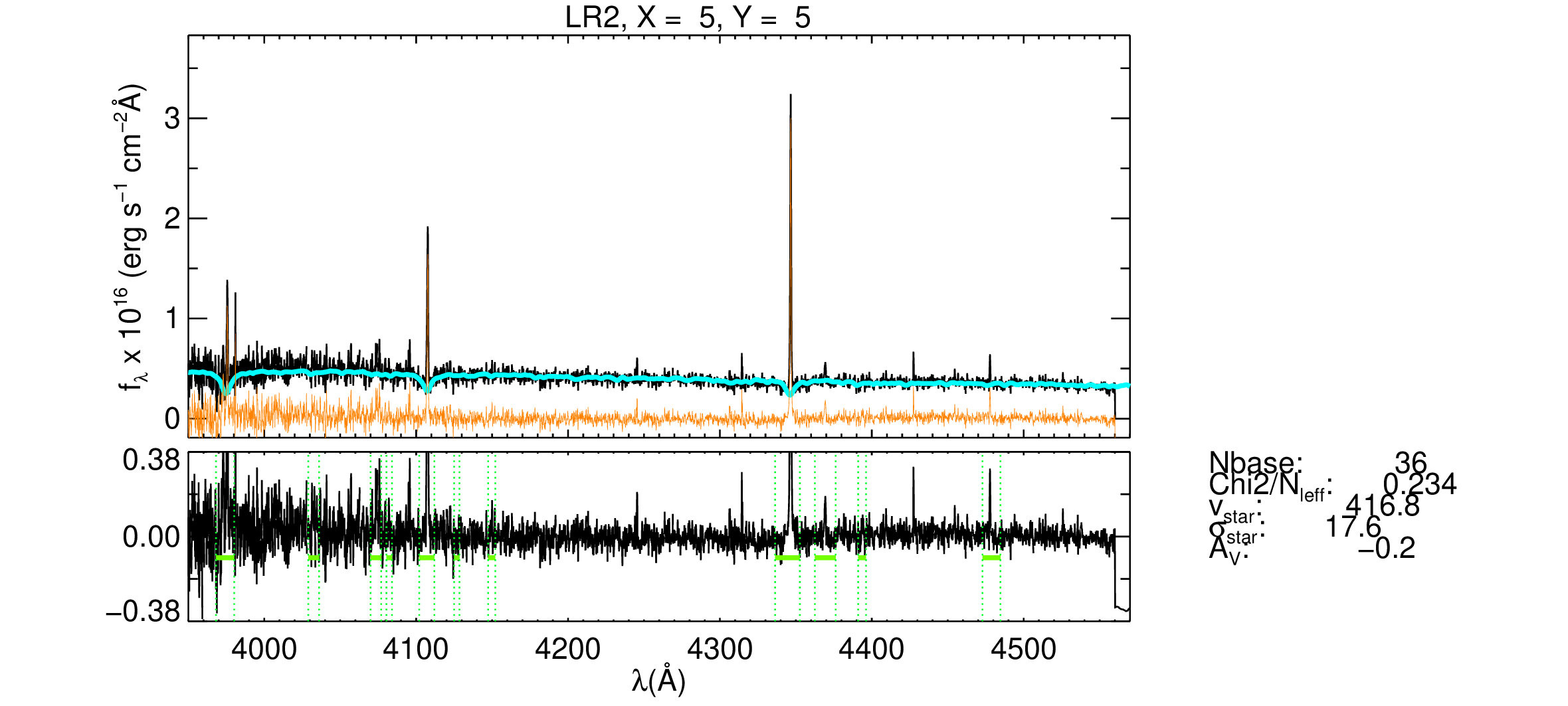}\\
\includegraphics[angle=0,width=0.45\textwidth, clip=,bb = 110 360 900 870]{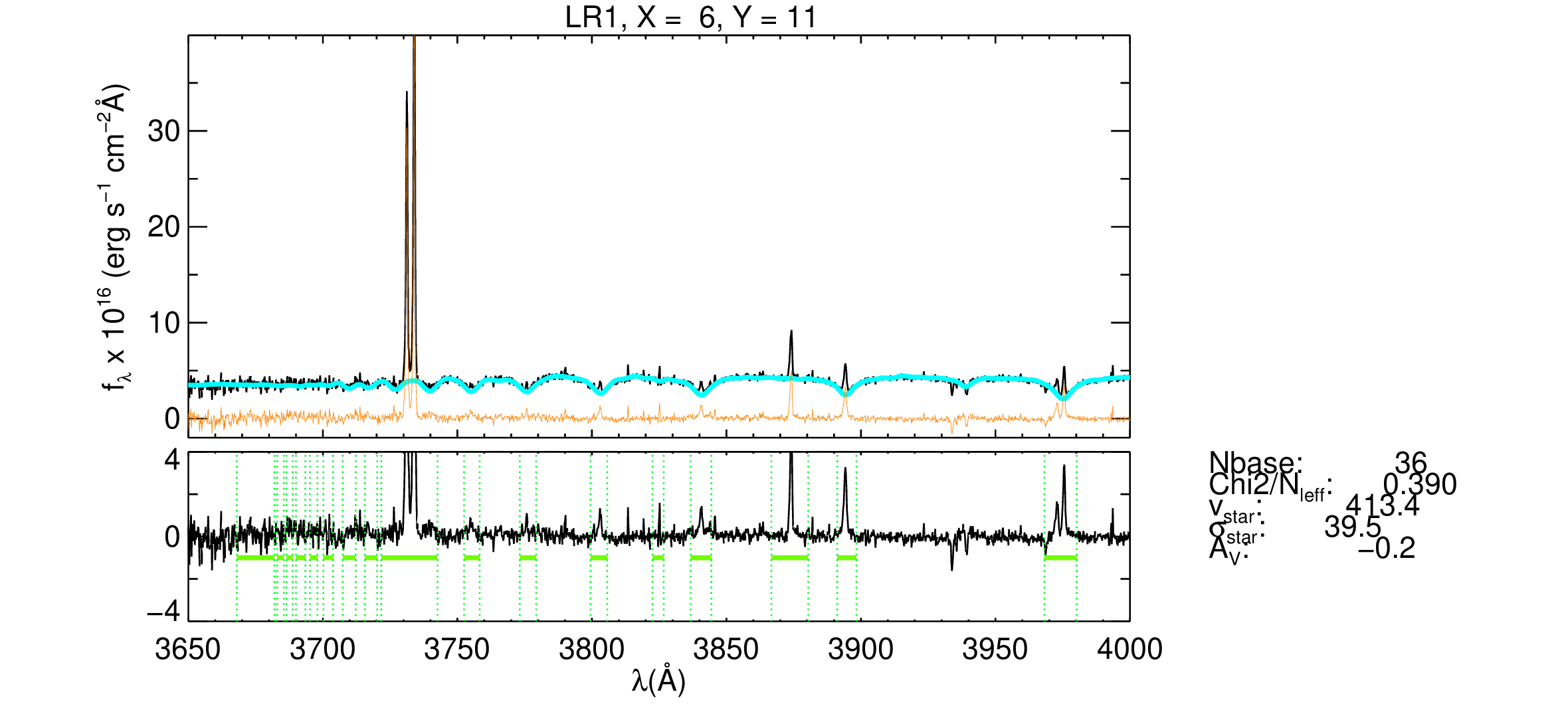}
\includegraphics[angle=0,width=0.45\textwidth, clip=,bb = 110 360 900 870]{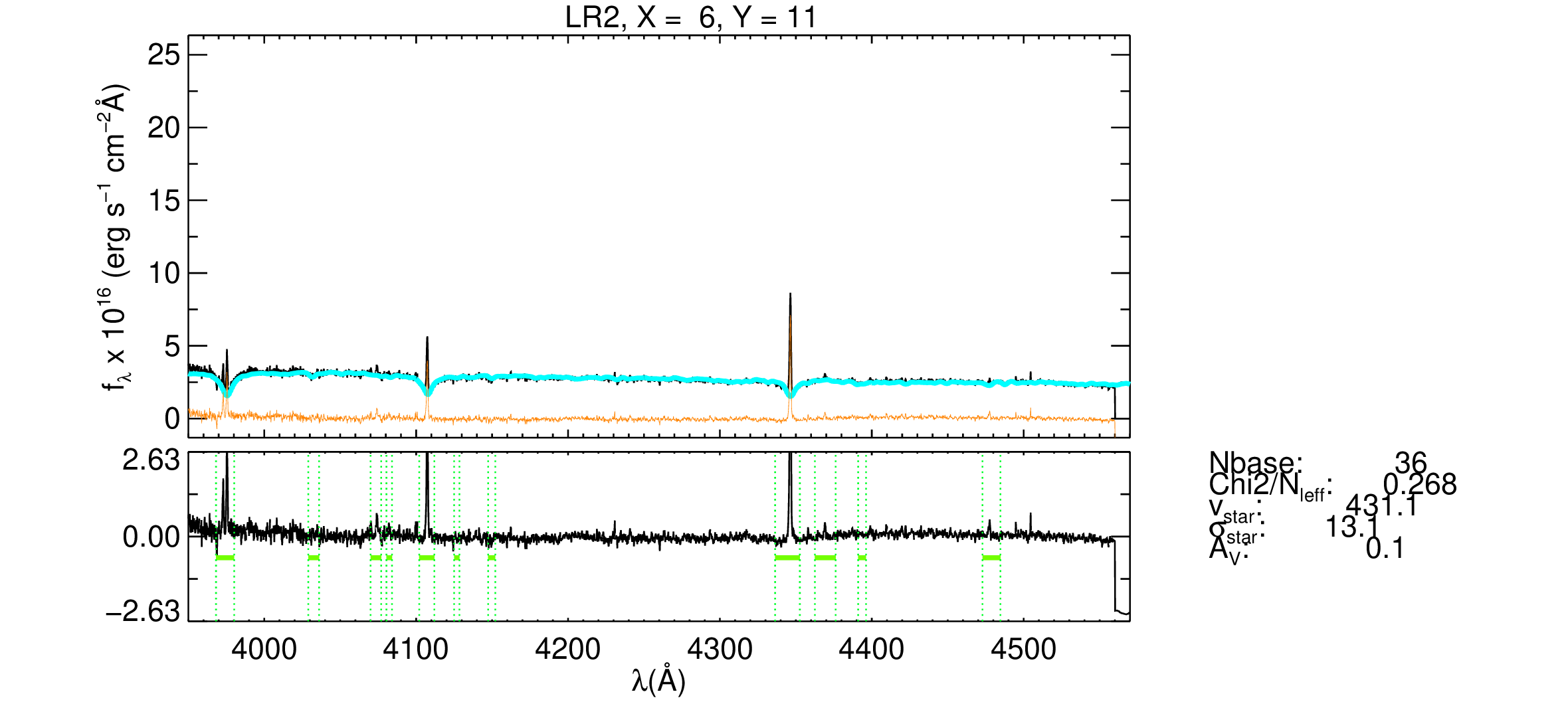}\\
\includegraphics[angle=0,width=0.45\textwidth, clip=,bb = 110 360 900 870]{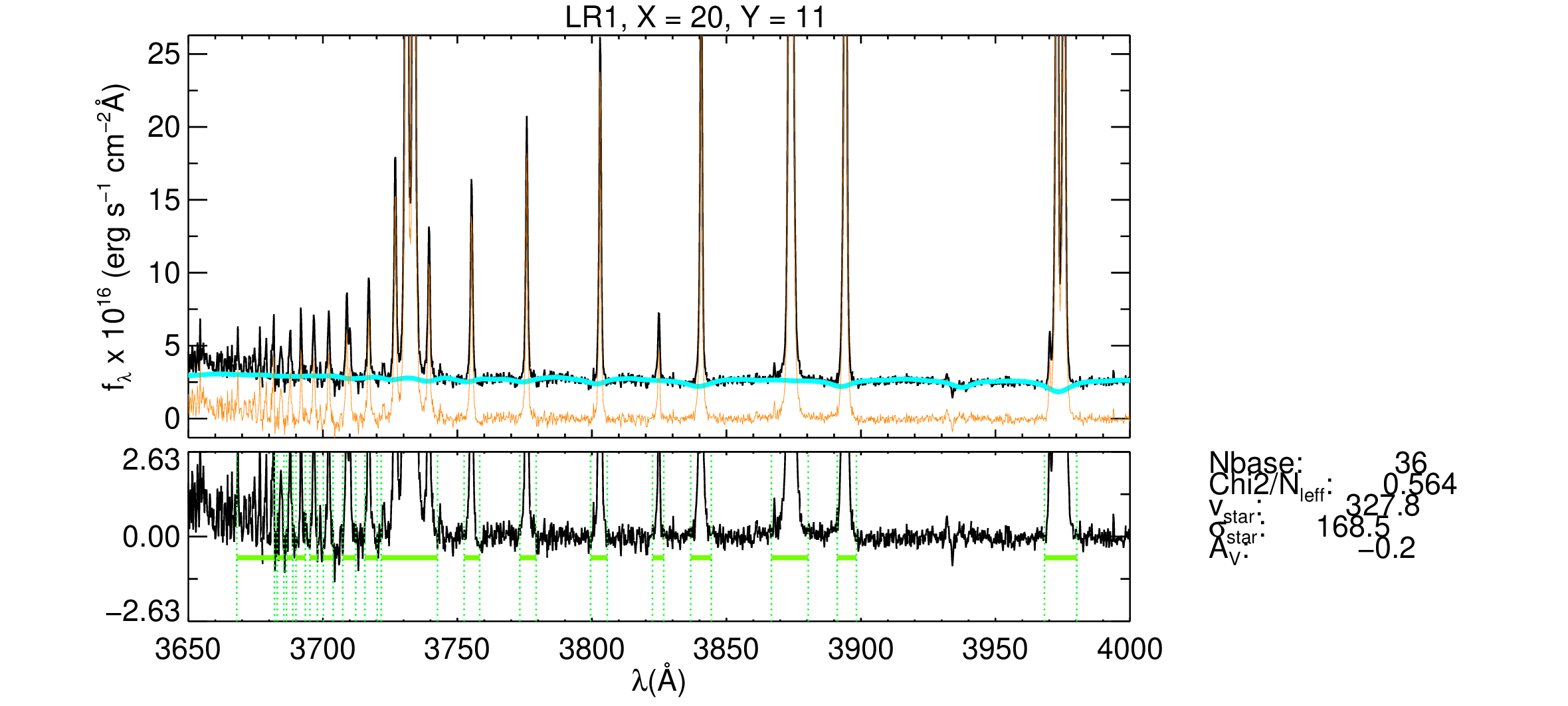}
\includegraphics[angle=0,width=0.45\textwidth, clip=,bb = 110 360 900 870]{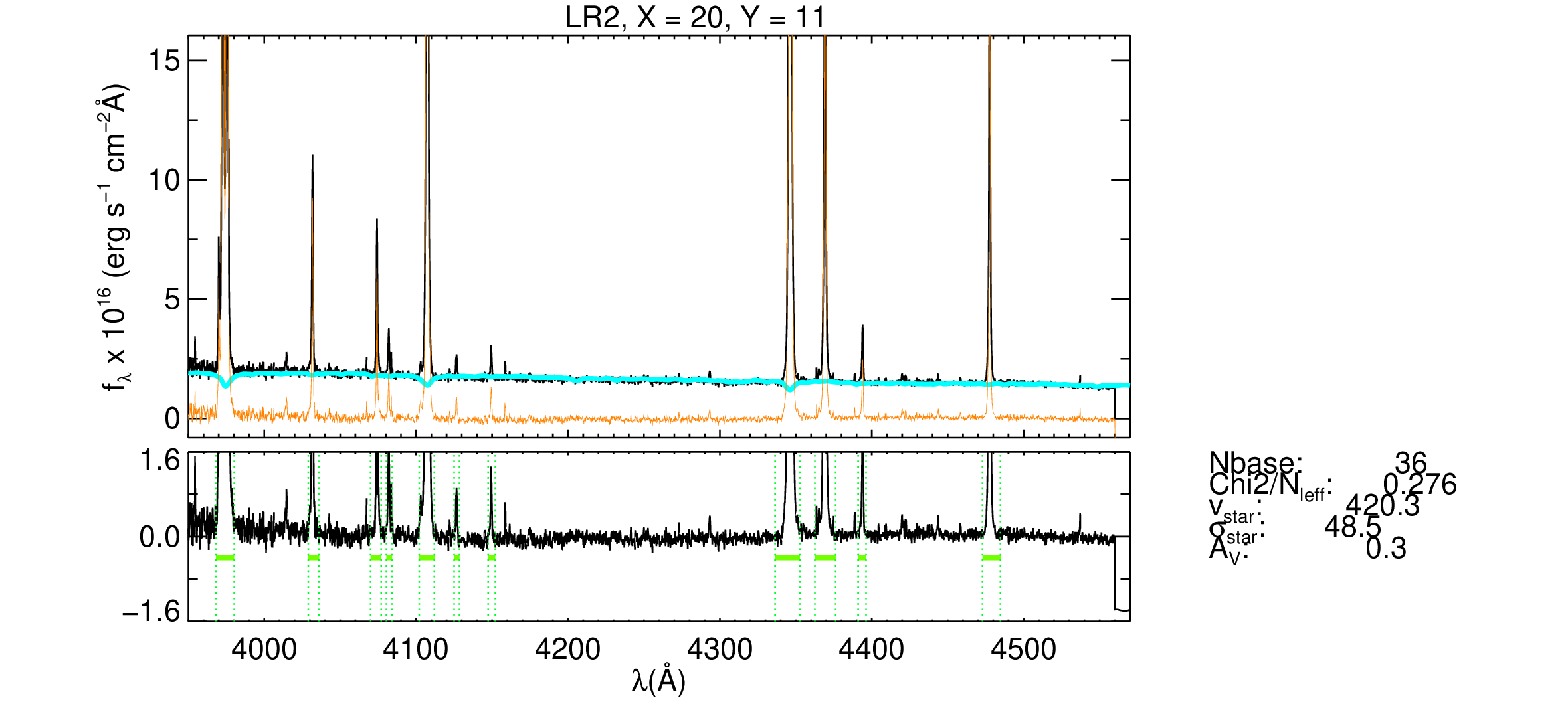}\\
   \caption[Representative spectra to illustrate the modelling and subtraction of the underlying stellar population emission.]{Representative spectra to illustrate the modelling and subtraction of the underlying stellar population emission for the LR1 (\emph{left}) and LR2 (\emph{right}) data cubes. The selected spaxels show typical emission in the lowest surface brightness area (\emph{first row}), the peak of continuum emission $\sharp$3, associated to two relatively old clusters (\emph{second row}), and the peak of emission in H$\gamma$ (\emph{third row}). Each individual graphic contains the observed, modelled and subtracted spectrum in black, cyan and orange, respectively. Below this main graphic there is an auxiliary one showing a zoom of the subtracted spectrum in black together with the spectral regions that were masked from the continuum fitting (green). \label{stellaremissionsubs}} 
 \end{figure*}

\subsection{Data reduction}

The processing of the data on \object{NGC~5253} for run 078.B-0043 is described in
\citetalias{mon10}. Those for run 383.B-0043 were processed using a combination
of the pipeline provided by ESO (version
2.8.9)\footnote{http://www.eso.org/projects/dfs/dfs-shared/web/vlt/vlt-instrument-pipelines.html.}
via \texttt{esorex}, version 3.9.0 and some IRAF\footnote{The Image
  Reduction and Analysis Facility \emph{IRAF} is 
  distributed by the National Optical Astronomy Observatories which is
  operated by the association of Universities for Research in
  Astronomy, Inc. under cooperative agreement with the National
  Science Foundation.} routines.

The corresponding master calibration files were
created with the ESO pipeline tasks \texttt{gimasterbias},
\texttt{gimasterflat} and \texttt{giwavecalibration}. In particular,
fifteen bias frames were used for the masterbias creation while six
and three continuum lamp exposures were needed for the creation of the
L385.7 and L427.2 masterflats, respectively.
In order to remove the cosmic rays in the target exposures, we combined all
the individual frames for a given grating using \texttt{imcombine}
with IRAF. After that, each combined frame was processed
using the ESO pipeline in 
order to perform bias subtraction, spectral tracing and
extraction, wavelength calibration and correction of fibre
transmission and data cube creation.

In both settings, four isolated and evenly distributed arc lines were
fitted by a Gaussian in each spectrum in order to estimate the uncertainties
in the wavelength calibration as well as the instrumental
width. Centroids of the lines were determined with an accuracy of
$\sim0.005$~\AA\  which, for the covered spectral range, translates
into velocities of $\sim$0.3-0.4~km~s$^{-1}$. We measured a spectral
resolution of FWHM $\sim$0.329~\AA\ and 0.446~\AA\ for the L385.7 and
L427.2, respectively. This translates to $\sigma_{instru} \sim$
10.9 and 13.3~km~s$^{-1}$.


In the next step, the sky background was subtracted. For that purpose, we created a good
signal-to-noise (S/N) spectrum by averaging the spectra of the sky
fibres in each combined frame which was subtracted from every spectrum. Fibers 
suffering significant contamination due to cross-talk of adjacent object fibres were 
excluded from this combination.

Relative flux calibration was performed within IRAF. The sky in the 
standard star frames was subtracted following the
same methodology as with the science frames and then 
a spectrum was formed for each calibration star by co-adding all the
fibres of each standard star frame. Then, a sensitivity function was
determined with the IRAF tasks \texttt{standard} and \texttt{sensfunc}
and science frames were calibrated with \texttt{calibrate}. 

Finally, since data were taken in clear conditions, it was necessary to place both datacubes on the same 
(relative) flux scale. We scaled each spectrum in the LR1 mode to the same level as that of LR2 using the H7 and [Ne\,\textsc{iii}]$\lambda$3967 emission lines, which were present in both instrumental set-ups.


\subsection{Subtracting the emission of the stellar population}

The spectral region analyzed in this work is rich in spectral features caused by the underlying stellar population (e.g. Balmer absorption lines).
Their effect on the estimation of the flux in the emission lines for the gas is negligible for the brightest lines and/or those in the area of the bright \ghiir. However, they may affect the measurements in the area of lower surface brightness and, especially in those spaxels associated with the relatively older clusters ($\sim$10~Myr)\footnote{Ages for the clusters associated to this knot were incorrectly quoted in our previous work. The ages estimated by \citet{har04}, for their clusters 3 and 5 were 8 and 11~Myr, respectively.} associated to the peak of emission in the continuum $\sharp$3 \citepalias[see ][for the definition of the peaks of emission]{mon10}.

To estimate and correct this effect, we modelled and subtracted out the contribution of the underlying stellar continuum using the STARLIGHT\footnote{http://starlight.ufsc.br/index.php?section=1} spectral synthesis code \citep{cid05, cid09}. This code reproduces a given observed spectrum by selecting a linear combination of a sub-set of $N_\star$ spectral components from a pre-defined set of base spectra. In our particular case, we utilized as base spectra a set of single star populations from \citet{bru03}. These are based on the Padova 2000 evolutionary tracks \citep{gir00} and assume Salpeter initial mass function between 0.1 and 100~M$_\odot$. Since \object{NGC~5253} has a metallicity of Z$\sim$0.3~Z$_\odot$\footnote{We assumed 12 + $\log$(O/H)$_\odot$ = 8.66 \citep{asp04}.} \citep[e.g.][]{kob99}, we utilized only base spectra with $Z = $0.004 and 0.008. For each metallicity, we selected a set of 18 spectra with ages ranging from 1~Myr to  2.5~Gyr. We allowed for a single extinction for all the base spectra that was modeled as a uniform dust screen with the extinction law by \citet{car89}.

The stellar spectral energy distribution at each spaxel was independently modelled for each cube. The spectral regions utilized in the fits were 3\,650-4\,060~\AA{} and the whole spectral range for the LR1 and LR2 gratings, respectively.
All the main outputs for each spaxel were then reformatted in order to create three cubes per grating with the total, gaseous and modelled stellar emission. Additionally, the information associated to auxiliary properties such as stellar extinction in the $V$ band, the stellar velocity and velocity dispersion, the reduced $\chi^2$ and the absolute deviation of the fit (in \%) were reformatted and saved as a file suitable for manipulation with standard astronomical software. Hereafter, we will use both terms \emph{map} and \emph{image} to refer to these kind of files.

Representative examples of the achieved subtraction of the stellar component are shown in Fig.~\ref{stellaremissionsubs} which contains the observed, modeled and subtracted spectra for spaxels in the lowest surface brightness area, at the peak of emission in the continuum $\sharp$3 and in the peak of emission for the ionized gas. The subtraction of the stellar continuum is satisfactory in all three cases. Note that given the spectral resolution of the FLAMES data (larger than the one of the base spectra), some fine structure (e.g. at $\sim$3\,930~\AA{}) cannot be reproduced. However, this does not affect the recovery of the gaseous emission 
in which we are primarily interested.  
 
%

   \begin{figure}[th!]
   \centering
\includegraphics[angle=0,width=0.48\textwidth,clip=,bb = 30 25 530 370]{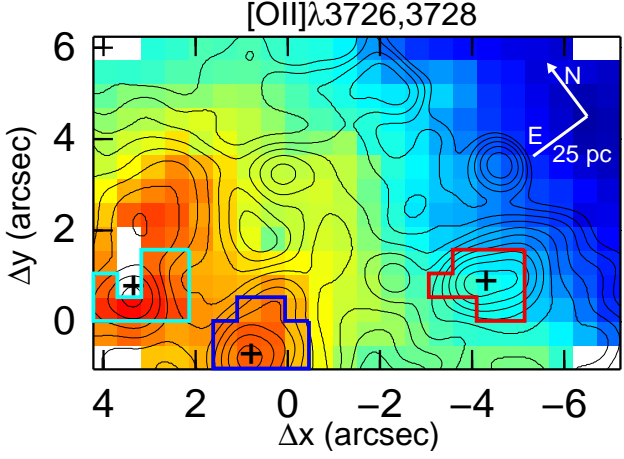}
   \caption[Covered field with FLAMES]{Ionized gas  distribution as traced by the \oii\ emission line doublet. We have over-plotted contours corresponding to the HST-ACS images in the F435W filter (programme 10609, P.I.: Vacca) convolved with a Gaussian of 0\farcs8 to simulate the seeing at Paranal. The position of the three main peaks of continuum emission are marked with crosses.
The map is presented in logarithmic scale in order to emphasize  the relevant morphological features and cover  a range of 1.4~dex.
Flux units are arbitrary. Utilized apertures to extract the selected spectra have been outlined according to the following color code: Cyan - Knot $\sharp$1; Blue - Knot $\sharp$2; Red - Knot $\sharp$3. Note the existence of three dead spaxels at $\sim[3\farcs.5,1\farcs0]$ as well as absence of signal in the spaxels at the four corners of the field of view. All of these spaxels will be marked hereafter by white rectangles. The scale and orientation are indicated.
 \label{estructura}}
 \end{figure}

\subsection{Line fitting and map creation}

As a first step, information from the relevant emission lines for each individual spaxel was obtained by fitting Gaussian functions in a semi-automatic way using the IDL based routine MPFITEXPR \citep{mar09}. Most of the lines (i.e. \hg-H8,  \oiii$\lambda$4363, \hei$\lambda$4009, \hei$\lambda$4026, \hei$\lambda$4388,  \hei$\lambda$4471, \sii$\lambda$4074 and \sii$\lambda$4076) were strong enough over most of the fov to be fitted independently. Exceptional cases were \neiii$\lambda$3967 and \hei$\lambda$3965 that were fitted simultaneously with H7, to assure a proper deblending of the emission lines and \oii$\lambda$3727 and \oii$\lambda$3729 that were fitted simultaneously.
For the rest of lines, the kinematic results for H7, present in both LR1 and LR2, were utilized as additional constrains. 
Several (strong) lines showed signs of asymmetries and/or multiple components in their profiles for a large number of spaxels. When possible, multi-component fits to these lines were also performed trying to assure a continuity in the derived physical (and chemical) properties between adjacent spaxels. In the present work, this multi-component fit will only be utilized in the discussion of the electron density structure.
Finally, as we did with the fitting of the underlying stellar population, we used the derived quantities together with the position within the data-cube for each spaxel to create a map. 

On account of the different pointings between the 078.B-0043 and 382.B-0043 observing runs, an important issue for this analysis was assuring a proper match between maps for quantities derived from the LR3 and LR6 cubes and those of the LR1 and LR2. As a first guess we used the three main peaks in the continuum images as our reference to align the images. However, given the relatively small fov of the IFU, the extended nature of knots $\sharp$1 and $\sharp$3 and that knot $\sharp$2 falls at the edge of the IFU in the LR1 and LR2 data, the derived offset were not accurate enough. To refine this first guess, we created maps for ratios involving lines observed with the two different pointings (e.g. \oiii$\lambda$5007/\oiii4363) using a grid of offsets centered in our first guess and in steps of 0.5~spaxels (=0\farcs26$\sim$1/4th seeing). 
We consider as final offset [2.5~spa,5.0~spa] (=[1\farcs3,2\farcs6]) which was the one providing the best continuity in adjacent spaxels for the variation of the line ratios. 
If the map of a given quantity was based on information from both runs, when possible, we independently carried out the calculations for 078.B-0043 and 383.B-0043. Only at the end, the offset was applied to the information associated to the 078.B-0043 run before carrying out the final calculation involving data from both runs. In this manner, the number of spatial interpolations was minimized.


\begin{table}
\small
\centering
     \caption[]{Observed fluxes for lines in the LR1 and the LR2 gratings with respect to F(H$\beta$).  \emph{(See text for details of the scaling of the spectra).}
\label{lineratios_lr1lr2}}
\scriptsize
            \begin{tabular}{ccccccccc}
            \hline
            \noalign{\smallskip}
 $\lambda_0$  & Ion & $f(\lambda)$ &   \multicolumn{3}{c}{F($\lambda$)/F(\hb)} \\
 (\AA)                &        &                         & Knot  $\sharp$1 & Knot $\sharp$2  & Knot $\sharp$3  \\
           \noalign{\smallskip}
\hline
            \noalign{\smallskip}
 3669.47 & \hi      & 0.268 &  0.17$\pm$0.05 &  0.42$\pm$0.05 & \ldots\\
 3671.48 & \hi      & 0.268 &  0.24$\pm$0.05 &  0.24$\pm$0.06 & \ldots\\
 3673.76 & \hi      & 0.268 &  0.35$\pm$0.05 &  0.46$\pm$0.06 & \ldots\\
 3676.37 & \hi      & 0.265 &  0.53$\pm$0.06 &  0.57$\pm$0.07 & \ldots\\
 3679.36 & \hi      & 0.265 &  0.52$\pm$0.04 &  0.69$\pm$0.11 & \ldots\\
 3682.81 & \hi      & 0.265 &  0.67$\pm$0.05 &  0.77$\pm$0.08 & \ldots\\
 3686.83 & \hi      & 0.263 &  0.75$\pm$0.05 &  0.99$\pm$0.08 & \ldots\\
 3691.56 & \hi      & 0.263 &  0.83$\pm$0.06 &  0.91$\pm$0.06 & \ldots\\
 3697.15 & \hi      & 0.261 &  1.10$\pm$0.08 &  0.91$\pm$0.06 & \ldots\\
 3703.86 & \hi      & 0.261 &  1.30$\pm$0.08 &  1.25$\pm$0.06 & \ldots\\
 3705.04 & \hei     & 0.261 &  0.70$\pm$0.07 &  0.57$\pm$0.06 & \ldots\\
 3711.97 & \hi      & 0.259 &  1.37$\pm$0.08 &  1.71$\pm$0.09 & \ldots\\
 3721.83 & \siii    & 0.257 &  2.95$\pm$0.16 &  2.68$\pm$0.09 & \ldots\\
 3726.03 & \oii     & 0.255 & 43.82$\pm$2.73 & 75.36$\pm$2.05 & 101.88$\pm$3.97 \\
 3728.82 & \oii     & 0.255 & 46.35$\pm$2.77 & 93.87$\pm$2.61 & 144.38$\pm$6.84 \\
 3734.17 & \hi      & 0.255 &  1.99$\pm$0.14 &  1.99$\pm$0.08 &   1.26$\pm$0.20 \\
 3750.15 & \hi      & 0.251 &  2.70$\pm$0.16 &  2.80$\pm$0.10 &   3.24$\pm$0.40 \\
 3770.63 & \hi      & 0.247 &  3.34$\pm$0.18 &  3.54$\pm$0.09 &   2.51$\pm$0.32 \\
 3797.90 & \hi      & 0.241 &  4.59$\pm$0.25 &  4.69$\pm$0.12 &   4.56$\pm$0.33 \\
 3819.61 & \hei     & 0.237 &  0.86$\pm$0.06 &  0.88$\pm$0.04 &   0.65$\pm$0.20 \\
 3835.39 & \hi      & 0.235 &  6.61$\pm$0.35 &  6.74$\pm$0.15 &   6.65$\pm$0.48 \\
 3868.75 & \neiii   & 0.227 & 48.87$\pm$3.10 & 31.73$\pm$0.93 &  22.34$\pm$0.89 \\
 3889.05 & \hi+\hei & 0.223 & 16.52$\pm$0.90 & 18.11$\pm$0.43 &  15.91$\pm$0.67 \\
 3964.73 & \hei     & 0.209 &  0.57$\pm$0.35 &  0.59$\pm$0.19 &   0.77$\pm$0.45 \\
 3967.46 & \neiii   & 0.207 & 15.88$\pm$0.82 & 10.38$\pm$0.31 &   8.77$\pm$0.59 \\
 3970.07 & \hi      & 0.207 & 15.80$\pm$0.82 & 15.80$\pm$0.38 &  15.80$\pm$0.72 \\
 4009.22 & \hei     & 0.198 &  0.20$\pm$0.03$^{\mathrm{(\ast)}}$ & \ldots & \ldots\\
 4026.21 & \hei     & 0.194 &  1.73$\pm$0.09 &  1.69$\pm$0.05 &   0.81$\pm$0.19 \\

            \noalign{\smallskip}
 \hline
            \noalign{\smallskip}
 4068.60 & \sii & 0.187      &      1.40$\pm$0.08   & 1.41$\pm$0.05  & 3.28$\pm$0.24 \\ 
 4076.35 & \sii & 0.184      &      0.46$\pm$0.03   & 0.52$\pm$0.03  & 1.25$\pm$0.16 \\ 
 4097.26 & \oiiB & 0.180     &      0.05$\pm$0.01   &  \ldots &      \ldots  \\ 
 4101.74 & \hi & 0.180       &     25.79$\pm$1.37   & 25.42$\pm$0.58  &	25.19$\pm$0.89  \\ 
 4120.82 & \hei & 0.176      &      0.20$\pm$0.02   &  0.09$\pm$0.03  & \ldots   \\ 
 4143.76 & \hei & 0.172      &      0.23$\pm$0.02   &  0.16$\pm$0.02  & \ldots   \\ 
 4243.97 & \feii & 0.153     &      0.03$\pm$0.01   &  0.05$\pm$0.01  & \ldots  \\ 
 4267.15 & \cii & 0.147      &      0.06$\pm$0.01   &  0.10$\pm$0.01  & \ldots  \\ 
 4287.40 & \feii & 0.144     &      0.12$\pm$0.01   &  0.24$\pm$0.02  &  1.23$\pm$0.15  \\ 
 4340.47 & \hi & 0.133       &     46.60$\pm$2.87   &  46.60$\pm$1.12 & 46.60$\pm$1.71  \\ 
 4359.34 & \feii & 0.129     &      0.06$\pm$0.01   &  0.13$\pm$0.01  &  0.53$\pm$0.10  \\ 
 4363.21 & \oiii & 0.129     &      7.77$\pm$0.39   &  3.70$\pm$0.08  &  2.64$\pm$0.21  \\ 
 4368.25 & \oi & 0.127       &      0.03$\pm$0.01   &  \ldots         &   \ldots  \\ 
 4387.93 & \hei & 0.122      &      0.49$\pm$0.03   &  0.45$\pm$0.02  &   \ldots  \\ 
 4413.78 & \feii & 0.118     &      0.07$\pm$0.01   &  0.01$\pm$0.01  &  \ldots\\ 
 4416.27 & \feii & 0.115     &      0.04$\pm$0.01   &  0.03$\pm$0.01  &  \ldots\\ 
 4437.55 & \hei & 0.110      &      4.48$\pm$0.25   &  \ldots  &      \ldots   \\ 
 4452.11 & \feii & 0.108     &      0.13$\pm$0.01   &  \ldots  &      \ldots  \\ 
 4471.48 & \hei & 0.103      &      0.08$\pm$0.01   &  0.19$\pm$0.01 &      0.36$\pm$0.18 \\ 
             \noalign{\smallskip}
           \hline
           \noalign{\smallskip}
\end{tabular}
\small
\begin{list}{}{}
\item[$^{\mathrm{(\ast)}}$] This line presented a  "secondary peak" towards the blue which correspond to a weak \feiii\ line at $\lambda$4008. At the resolution of the present data a proper deblending of the lines was not feasible.
\end{list}
\end{table}



\begin{table}
\small
\centering
\caption[]{Observed fluxes for lines in the LR3 and the LR6 gratings with respect to F(H$\beta$).  \emph{(See text for details of the scaling of the spectra).} \label{lineratios_lr3lr6}}
\scriptsize
            \begin{tabular}{ccccccccc}
            \hline
            \noalign{\smallskip}
$\lambda_0$  & Ion & $f(\lambda)$ &   \multicolumn{3}{c}{F($\lambda$)/F(\hb)} \\
 (\AA)                &        &                         & Knot  $\sharp$1 & Knot $\sharp$2  & Knot $\sharp$3 \\
\hline
4562.60 & \mgi & 0.080    &  0.11$\pm$0.01  &	0.19$\pm$0.01 	 &0.86$\pm$     0.19   \\  
 4571.00 & \mgi & 0.078   &  0.12$\pm$0.01  &	0.16$\pm$0.01 	 &0.74$\pm$     0.17   \\  
 4649.13 & \oiiB & 0.057  &  0.08$\pm$0.01  &  \ldots &    \ldots   \\  
 4658.10 & \feiii & 0.054 &  0.96$\pm$0.05  &	0.80$\pm$0.03 	 &2.34$\pm$     0.19   \\  
 4701.53 & \feiii & 0.043 &  0.27$\pm$0.01  &	0.23$\pm$0.01 	 &0.84$\pm$     0.19   \\  
 4711.37 & \ariv & 0.041  &  1.17$\pm$0.06  &	0.19$\pm$0.00 	 &0.57$\pm$     0.02   \\  
 4713.14 & \hei & 0.041   &  0.65$\pm$0.05  &	0.41$\pm$0.01 	 &0.19$\pm$     0.16   \\  
 4740.16 & \ariv & 0.032  &  1.18$\pm$0.05  &	0.17$\pm$0.01 	 & \ldots   \\  
 4754.83 & \feiii & 0.030 &  0.19$\pm$0.01  &	0.18$\pm$0.01 	 & \ldots   \\  
 4769.60 & \feiii & 0.024 &  0.10$\pm$0.01  &	0.10$\pm$0.01 	 & \ldots   \\  
 4814.55 & \feii & 0.013  &  0.06$\pm$0.01  &	      \ldots  &      \ldots   \\ 
 4861.33 & \hi & 0.000    &  100.00$\pm$4.36  &    100.00$\pm$1.82     &100.00$\pm$     2.51   \\  
 4881.00 & \feiii & $-$0.005 & 0.37$\pm$0.02  &	0.25$\pm$0.02 	 & \ldots  \\  
 4921.93 & \hei & $-$0.016   & 1.14$\pm$0.05  &	1.03$\pm$0.03 	 & \ldots  \\  
 4931.32 & \oiii & $-$0.019  & 0.16$\pm$0.01  &	0.16$\pm$0.02 	 & \ldots  \\  
 4958.91 & \oiii & $-$0.027  & 232.56$\pm$10.88  &    162.83$\pm$3.25  &   114.29$\pm$     3.25  \\  
 4985.90 & \feiii & $-$0.032 & 0.47$\pm$0.03  &	0.64$\pm$0.03 	 &2.78$\pm$0.26  \\  
 5006.84 & \oiii & $-$0.040  & 732.56$\pm$34.05  &    508.85$\pm$10.23  &   358.10$\pm$10.25  \\  
 5015.68 & \hei & $-$0.043   & 1.87$\pm$0.09  &	2.19$\pm$0.06 	 &2.11$\pm$0.25  \\  
 5041.03 & \siiiB & $-$0.049 & 0.24$\pm$0.02  &	0.16$\pm$0.01 	& \ldots   \\ 
 5055.98 & \siiiB & $-$0.055 & \ldots  &	0.16$\pm$0.01 	& \ldots   \\  
 \hline
 6548.03 & \nii & $-$0.312   & 10.70$\pm$0.46   & 5.88$\pm$0.16 &     11.33$\pm$     0.29      \\ 
 6562.82 & \hi & $-$0.314    & 400.00$\pm$16.86   & 391.15$\pm$7.37 &    415.24$\pm$    10.18  \\ 
 6583.41 & \nii & $-$0.316   & 32.67$\pm$1.39   &  17.96$\pm$0.49 &     35.05$\pm$     0.80  \\ 
 6678.15 & \hei & $-$0.329   & 4.55$\pm$0.20   &   4.20$\pm$0.08 &     \ldots  \\ 
 6716.47 & \sii & $-$0.332   & 13.95$\pm$0.60   &  20.53$\pm$0.41 &     55.24$\pm$1.32  \\ 
 6730.85 & \sii & $-$0.335   & 12.91$\pm$0.55   &  16.55$\pm$0.34 &     40.76$\pm$0.98  \\ 
 7002.23 & \oi & $-$0.365    & 0.09$\pm$0.01   &   0.10$\pm$0.01 &     \ldots  \\ 
 7065.28 & \hei & $-$0.375   & 7.73$\pm$0.30   &   3.85$\pm$0.08 &      2.83$\pm$0.19   \\ 
 7135.78 & \ariii & $-$0.384 & 16.34$\pm$0.73   &  12.65$\pm$0.26 &     10.67$\pm$0.30  \\ 
\noalign{\smallskip}  
\hline
\end{tabular}
\end{table}


\begin{table}
\small
     \centering
     \caption[]{Physical properties of the ionized gas. \label{physiprop}}
            \begin{tabular}{lcccccccc}
            \hline
            \noalign{\smallskip}
                     & Knot  $\sharp$1 & Knot $\sharp$2  & Knot $\sharp$3  \\
\hline
\noalign{\smallskip}  
 $E(B-V)$ & 0.31$\pm$0.04 & 0.29$\pm$0.02 & 0.35$\pm$0.04 \\
\noalign{\smallskip}  
\hline
\noalign{\smallskip}  
$n_e$(cm$^{-3}$) (\oii) & 310$^{+150}_{-140}$  & 140$^{+50}_{-50}$  & 30$^{+100}_{-60}$  \\
$n_e$(cm$^{-3}$) (\sii) & 385$^{+150}_{-120}$  & 170$^{+35}_{-30}$  & 70$^{+30}_{-25}$  \\
$n_e$(cm$^{-3}$) (\ariv) & 4\,600$^{+1200}_{-1000}$ & 2\,800$^{+1800}_{-900}$ & \ldots \\
$n_e$(cm$^{-3}$) (\feiii) & 270$^{+40}_{-30}$ & 155$^{+20}_{-15}$ & 95$^{+25}_{-20}$ \\
\noalign{\smallskip}  
\hline
\noalign{\smallskip}  
$T_e$ (K) (\oiii) & 11\,570$^{+420}_{-350}$ &  10\,280$^{+130}_{-120}$ & 10\,320$^{+350}_{-300}$  \\
$T_e$ (K) (\oii)  & 11\,360 &  10\,470 & 10\,500 \\
$T_e$ (K) (\sii) & 9\,700$^{+1700}_{-1300}$ & 8\,700$^{+490}_{-440}$ & 8\,900$^{+870}_{-680}$ \\
\noalign{\smallskip}  
\hline
\end{tabular}
\begin{flushleft}
\begin{itemize}
\item Ratios for $n_e$: \oii: \oii$\lambda$3727/\oii$\lambda$3729; \sii: \sii$\lambda$6717/\sii$\lambda$6731; \ariv: \ariv$\lambda$4711/\ariv$\lambda$4740; \feiii: \feiii$\lambda$4986/\feiii$\lambda$4658\\
\item Ratios for $T_e$: \oiii: \oiii$\lambda\lambda$4959,5007/\oiii$\lambda$4363; \sii: \sii$\lambda\lambda$6716,6731/\sii$\lambda$4069,4076. Note that $T_e$(\oii) was not obtained directly from measurements of \oii\ lines. Instead, it was derived from $T_e$(\oiii) and $n_e$(\oii) according to the expressions provided in \citet{gar09}.\\
\end{itemize}
\end{flushleft} 
\end{table}

\begin{table}
     \centering
\small
     \caption[]{Ionic chemical abundances of the ionized gas. \label{chemiprop}}
            \begin{tabular}{lcccccccc}
            \hline
            \noalign{\smallskip}
                     & Knot  $\sharp$1 & Knot $\sharp$2  & Knot $\sharp$3  \\
            \noalign{\smallskip}
\hline
            \noalign{\smallskip}
$12+\log(O^+/H^+)$    & 7.33$\pm$0.10 & 7.73$\pm$0.04 & 7.88$\pm$0.09 \\
$12+\log(O^{2+}/H^+)$ & 8.18$\pm$0.09 & 8.17$\pm$0.03 & 8.03$\pm$0.07 \\
$12+\log(O/H)$        & 8.24$\pm$0.09 & 8.32$\pm$0.04 & 8.26$\pm$0.08 \\
            \noalign{\smallskip}
\hline
            \noalign{\smallskip}
$12+\log(S^+/H^+)$    &  5.71$\pm$0.07   & 5.96$\pm$0.03 &  6.31$\pm$0.05 \\
$\log(S^+/O^+)$    & $-$1.62$\pm$0.17 &  $-$1.77$\pm$0.07 &  $-$1.57$\pm$0.14 \\
            \noalign{\smallskip}
\hline
            \noalign{\smallskip}
$12+\log(N^+/H^+)$    & 6.52$\pm$0.07 & 6.36$\pm$0.03 & 6.62$\pm$0.06 \\
$12+\log(N/H)$        & 7.43$\pm$0.07 & 6.94$\pm$0.03 & 7.00$\pm$0.06 \\ 
ICF ($N^{+}$)                         & 8.04 & 3.84 & 2.40\\
$\log(N/O)$  	       & $-$0.81$\pm$0.25 & $-$1.37$\pm$0.11 & $-$1.26$\pm$0.22 \\  
            \noalign{\smallskip}
\hline
            \noalign{\smallskip}
$12+\log(Ne^{2+}/H^+)$    & 7.49$\pm$0.10 & 7.50$\pm$0.04 & 7.34$\pm$0.09 \\
ICF (Ne)  & 1.14  & 1.35  & 1.71\\
$12+\log(Ne/H)$    & 7.55$\pm$0.10 & 7.63$\pm$0.04 & 7.57$\pm$0.09  \\
$\log(Ne/O)$  & $-$0.69$\pm$0.28 & $-$0.69$\pm$0.12 & $-$0.69$\pm$0.25 \\  
            \noalign{\smallskip}
\hline
            \noalign{\smallskip}
$12+\log(Ar^{2+}/H^+)$    &  5.89$\pm$0.07 &  5.90$\pm$0.03 & 5.80$\pm$0.05 \\
$12+\log(Ar^{3+}/H^+)$    &  4.84$\pm$0.08 &  4.15$\pm$0.05 & \ldots        \\
ICF  (Ar)$^{(\ast)}$      &  1.02$\pm$0.04 &  1.07$\pm$0.05 & 1.45$\pm$1.91 \\
$12+\log(Ar/H)$           &  5.90$\pm$0.22 &  5.93$\pm$0.12 & 5.96$\pm$1.45 \\
$\log(Ar/O)$              &  $-$2.34$\pm$0.30   & $-$2.38$\pm$0.16  & $-$2.30$\pm$1.53 \\  
            \noalign{\smallskip}
\hline
            \noalign{\smallskip}
$12+\log(Fe^{2+}/H^+)$    & 5.58$\pm$0.09  & 5.66$\pm$0.04 &  6.12$\pm$0.09 \\
\noalign{\smallskip}  
\hline
\end{tabular}
\begin{list}{}{}
\item $^{(\ast)}$ ICF  (Ar$^{2+}$) for knot $\sharp$3.
\end{list} 
\end{table}

\section{Results \label{resultados}}

\subsection{Gaseous emission in selected apertures \label{selaper}}

The main aim of the present work is to provide a full 2D characterization of the physical and chemical properties of the ionized gas in the central region of the galaxy. However, there are reasons to perform a more detailed analysis in specific regions. Firstly, we can define very precisely the aperture utilized to extract a given spectrum, avoiding the limitations associated to the positioning of a slit and assuring that the spectrum is spatially associated with a given feature of interest (e.g. the location of a star cluster). As an example,  \citet{lop07} characterized the properties of our knot $\sharp$2 (their UV-1). However, neither their positions HII-1 nor HII-2 sample the gas associated to the peak of emission in \ha, but some emission associated to the \ghiir\ ~towards the west and south of the peak of emission. Secondly, a comparison of the measurements provided here, with those already existing in the literature serves as a sanity check which will reinforce the results obtained in the 2D mapping.
Thirdly, most of the present methodologies used in the analysis of the physical and chemical properties of the ionized gas were developed to study \emph{complete} H\,\textsc{ii} regions. At present, the astronomical community is still in the process of understanding under which conditions these can be applied to small portions of them  \citep[e.g.][]{per11,erc12}. The areas analyzed in this section will be closer to this notion of \emph{complete} H\,\textsc{ii} than individual spaxels.
Finally, by combining the signal of a collection of spaxels, a larger signal-to-noise (S/N) ratio is achieved and fainter lines can be detected. 

As we saw in Paper I, the stellar emission is dominated by three peaks of emission named as knots $\sharp$1, $\sharp$2 and $\sharp$3 (see Tab. 3 of Paper~I to establish the correspondence between this nomenclature and those of previous studies). Note that knot $\sharp$1 is associated to the main \ghiir\ marked in Fig. \ref{apuntado}. The spectra associated to them have been co-added and extracted in order to determine relative line intensities in their surrounding ionized gas. 
Utilized apertures in the LR1 and LR2 data are outlined in Fig. \ref{estructura}.  For the LR3 and LR6, we applied an offset of 2 and 5 spaxels in the \emph{x} and \emph{y} direction respectively. Since the offset between the two pointings was determined as [2.5~spa,5.0~spa], this implies a small difference in the selected apertures from the old and the new data which introduces an extra uncertainty when comparing fluxes of lines belonging to different set of data. In consequence, the line fluxes for the LR3 and the LR6 gratings were directly measured with respect to \hb\ and \ha\ while those for the LR1 and LR2 gratings where measured with respect to \hg\ and H7, and then
we assumed the theorietical Balmer line intensities obtained from \citet{sto95} for Case~ B, $T_e=10^4$~K and $n_e=100$~ cm$^{-3}$.
%
Most of the lines were measured independently by fitting the emission to a single Gaussian profile using \texttt{mpfit}. Exceptionally, lines in the pairs H16 and \hei$\lambda$3705, \feii$\lambda$4414 and \feii$\lambda$4416, and \ariv$\lambda$4711 and \hei$\lambda$4713 were simultaneously fitted due to their proximity in wavelength. Similarly, all the three lines \hei$\lambda$3965, \neiii$\lambda$3967 and H7 were fitted at once.
Measurements for all detected lines are compiled in Tables \ref{lineratios_lr1lr2} and \ref{lineratios_lr3lr6}.

\subsubsection{Physical conditions of the ionized gas \label{fiscon}}

Derived physical conditions for the ionized gas are listed in Table \ref{physiprop}.
The values of reddening were derived from the measured \ha/\hb\ line ratio following the methodology described in Paper~I.
Electron temperature ($T_e$) and density ($n_e$) based on \sii, \oii\ and \oiii\ emission lines (and thus tracing different layers in the ionization structure) were derived using the expressions provided by \citet{gar09}\footnote{http://www.iac.es/consolider-ingenio-gtc/index.php?option=com\_content\&view=article\&id=223:qa-spatially-resolved-study-of-ionized-regions-in-galaxies-at-different-scales\&catid=45:tesis\&Itemid=65}.
For that we proceeded as follows: we assumed an initial $n_e$ of 100~cm$^{-3}$ to obtain a first guess of the different $T_e$'s. The resulting $T_e$(\oii) and $T_e$(\sii) were used as input to obtain new estimates of $n_e$(\oii) and $n_e$(\sii) and the process was iterated until convergence. Typically one or two iterations were enough. 
These functions reproduce the predictions of the task \texttt{temden}, based on the \texttt{fivel} program \citep{sha95} included in the IRAF package \texttt{nebular}.
They used the same  atomic coefficients as in \citet{per03}, except for $O^+$ for which they used the transition
probabilities from \citet{zei82} and the collision strengths from \citet{pra76}.
$T_e$(\oii) was not independently derived from our data, since the \oii$\lambda\lambda$7320,7330 doublet was not covered in these observations. Instead, we derived $T_e$(\oii) from $T_e$(\oiii) and $n_e$(\oii) using the models presented by \citet{per03}. As an improvement to former relations based on modelling \citep[e.g.][]{sta90}, these take into account dependencies on $n_e$. At low densities, for a given $T_e$(\oiii), models by \citet{per03} predict only slightly (i.e. $\lsim2$\%) lower temperatures than those by \citet{sta90} while in the densest zones of NGC~5253 (e.g. our knot $\sharp$1), estimated $T_e$(\oii) can be up to $\sim$17\% smaller.
For $n_e$(\feiii), we used the line ratios tabulated by \citet{kee01}, assuming $T_e=12\,000$~K. Note that for the utilized emission lines, the dependence of the derived $n_e$ on the assumed temperature is negligible.
Finally, $n_e$(\ariv) was derived using directly \texttt{temden} and assuming $T_e$(\ariv)=$T_e$(\oiii).

Derived $T_e$ and $n_e$ for our knots $\sharp$1 and $\sharp$2 agree well within the uncertainties with the values reported by \citet{sid09} for his apertures A and B, respectively. Additional measurements for $T_e$ and $n_e$ in knot $\sharp$2 are provided by \citet{lop07}, \citet{kob97} and \citet{gus11}. While our derived $T_e$'s are in relatively good agreement, with $T_e$(\oiii) and $T_e$(\sii) being slightly lower and larger respectively than the values reported in these works, there is a discrepancy between the different measurements of $n_e$ in the literature. A comparison of our $n_e$(\sii) for knot $\sharp$ 1 and the values reported by \citet{gus11} for their apertures C1 and P2 shows a discrepancy between the reported values. Since $n_e$ varies a lot in this area (a factor of $\sim$2-3 on scales of $\sim$0\farcs5, see Fig. \ref{mapne} in Sec. \ref{densistruc}), the precise definition of the aperture (i.e. size, position and shape) in each case seems the most plausible explanation for this discrepancy.
 Our $n_e\sim150$~cm$^{-3}$ in knot $\sharp$2 is consistent within the errors with being below the low density limit as reported by \citet{kob97}. However, it is a factor $\sim$2 lower than the values reported by \citet{lop07}. While estimated uncertainties could account for the differences in $n_e$(\sii), one must resort to other causes, such as differences in the definition of the aperture, to explain the discrepancies in $n_e$(\oii).

\subsubsection{Chemical abundances \label{abunaper}}

Abundances for the different ionic species are listed in Table \ref{chemiprop} and were derived using the relations provided in Appendix~B.2 of \citet{gar09}. These are appropriate fittings to the results of the IRAF task \texttt{ionic} \citep {sha95}, based on the 5-level atom program developed by \citet{der87}, and follow the functional form given by \citet{pag92}.
For the abundances of neon, argon, iron and O$^{++}$, we utilized $T_e$(\oiii). For O$^{+}$ and nitrogen, we utilized $T_e$(\oii), derived from $T_e$(\oiii) and $n_e$(\oii). Finally, for sulfur, we took $T_e$(\sii). Differences in abundances due to the assumed $T_e$ are discussed below.
To derive the total abundances, unseen ionization stages of a given element were taken into account by including the appropriate ionization factor (ICF) for each species when necessary, following the prescriptions provided by \citet{kin94} for all the elements but argon. For this element, we utilized those provided by \citet{izo94}.

\emph{Oxygen:} We assumed $O/H = (O^+ + O^{2+})/H^+$. The derived values are in good agreement  with those provided in the literature using long-slit data \citep[e.g.][]{gus11,sid09,lop07,kob97}. The largest differences are for knot $\sharp$1 that shows abundances $\sim$0.1~dex higher than those measured by \citet{gus11}. This is however within the uncertainties.

\emph{Sulfur:} Derived values are typically $\sim$0.2~dex larger than those provided by \citet{lop07}. Differences can be attributed to the adopted $T_e$. \citet{lop07} use the average of $T_e$(\sii), $T_e$(\oii), and $T_e$(\nii) that can be $\sim$2\,000~K larger than $T_e$(\sii). If a $T_e$(\oii), derived from the $n_e$(\oii) and $T_e$(\oiii), were utilized, our abundances would become typically $\sim$0.15~dex smaller in agreement with the values reported by these authors. A similar effect is observed when comparing with the results reported by \citet{sid09} who utilized $T_e$(\oiii) to derive all the abundances and those reported by \citet{gus11} who used a $T_e$(\oii) $\sim$1\,500-2\,000~K larger than the $T_e$(\sii) utilized here.


\emph{Nitrogen:} We assumed $N/O=N^+/O^+$, which is an accurate approximation to about $\pm$20\% for nebulae with metallicities smaller than that in the LMC \citep{gar90}. 
In agreement with previous work, while knots $\sharp$2 and $\sharp$3 present relative abundances in nitrogen within the range expected for galaxies at this metallicity \citep[see e.g.][]{mol06}, the aperture associated to knot $\sharp$1  present a clear excess in nitrogen. Interestingly, although consistent with the $N/O$ expected for galaxies at this metallicity, knot $\sharp$3 has a slightly larger abundance than knot $\sharp$2.


\emph{Neon:} Since the ionization structure is similar to the one of the oxygen, we assumed $Ne^{2+}/Ne = O^{2+}/O$. Reported values agree within the errors with previous measurements in the literature \citep{gus11,lop07,kob97}.

\emph{Argon:} Total argon abundance was derived using the ICFs based on the expressions provided by \citet{izo94}. Derived ratios are within the range of those previously published \citep{gus11,lop07,sid09} and consistent with a homogeneous $Ar/O$ ratio across the face of the galaxy.

\emph{Iron:} Reported values for Fe$^2+$ agree with those given by \citet{lop07}.

To recap, abundances of the different ionic species in knot~$\sharp$1 and knot~$\sharp$2 agree in general with those previously reported \citep{kob97,lop07,sid09}. Abundances in knot~$\sharp$3 (not reported so far)  are similar to those in knot $\sharp$1 and knot $\sharp$2. With the exception of the relative abundance for $N/O$ in knot~$\sharp$1 and maybe knot~$\sharp$3, no chemical species seems overabundant in any of the selected apertures.
This is in agreement with the relative enrichment in nitrogen 	previously widely reported.


   \begin{figure}[th!]
   \centering
\includegraphics[angle=0,width=0.48\textwidth,clip=,bb = 30 110 530 415]{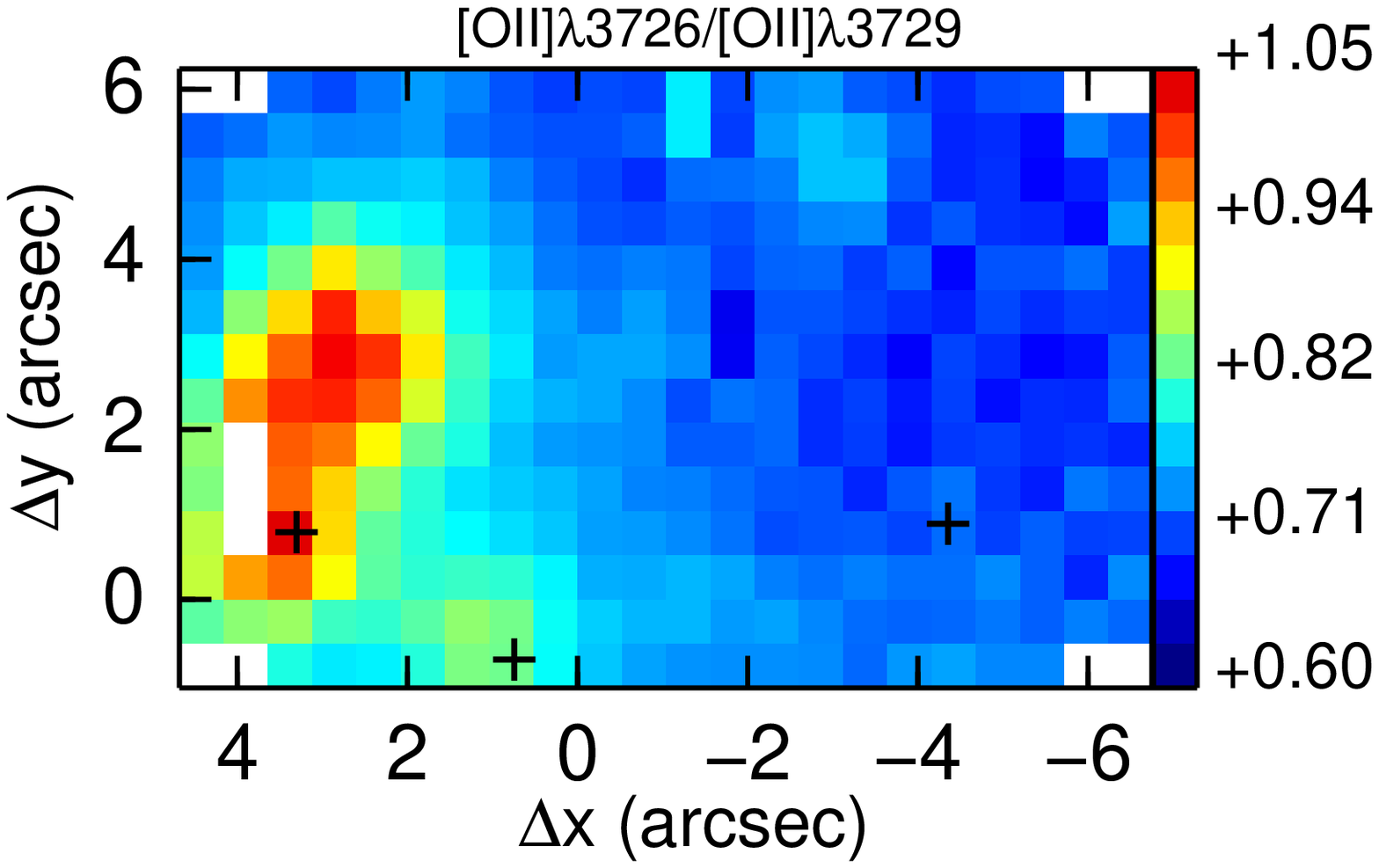}
\includegraphics[angle=0,width=0.48\textwidth,clip=,bb = 30 110 530 415]{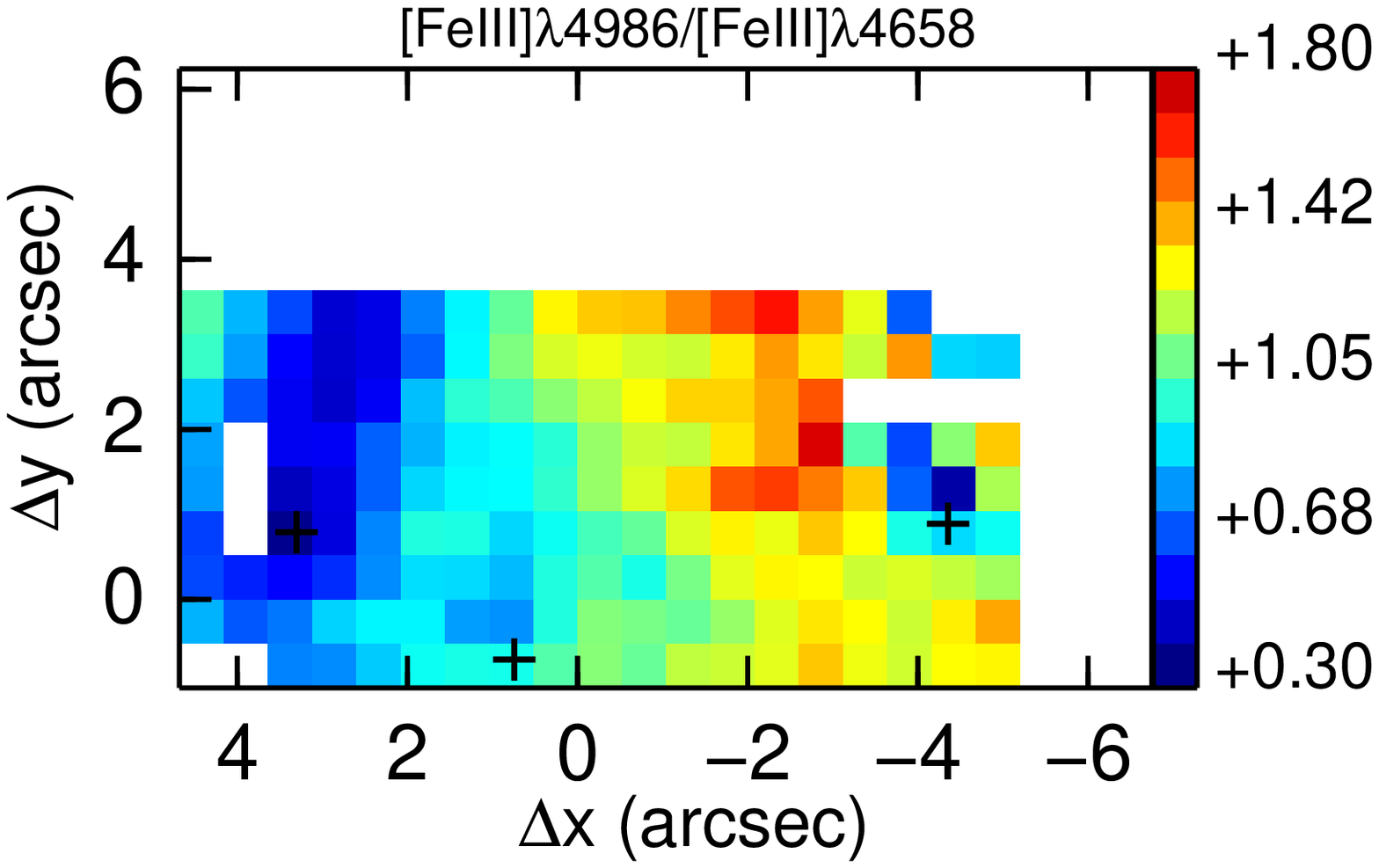}
   \caption[Maps for the two line ratios sensitive to the $n_e$]{Maps for the new line ratio sensitive to the electron density. The position of the three main peaks of continuum emission are shown as crosses for reference. 
 \label{mapo2rat}}
 \end{figure}

   \begin{figure}[th!]
   \centering
\includegraphics[angle=0,width=0.48\textwidth,clip=, bb = 30 110 530 415]{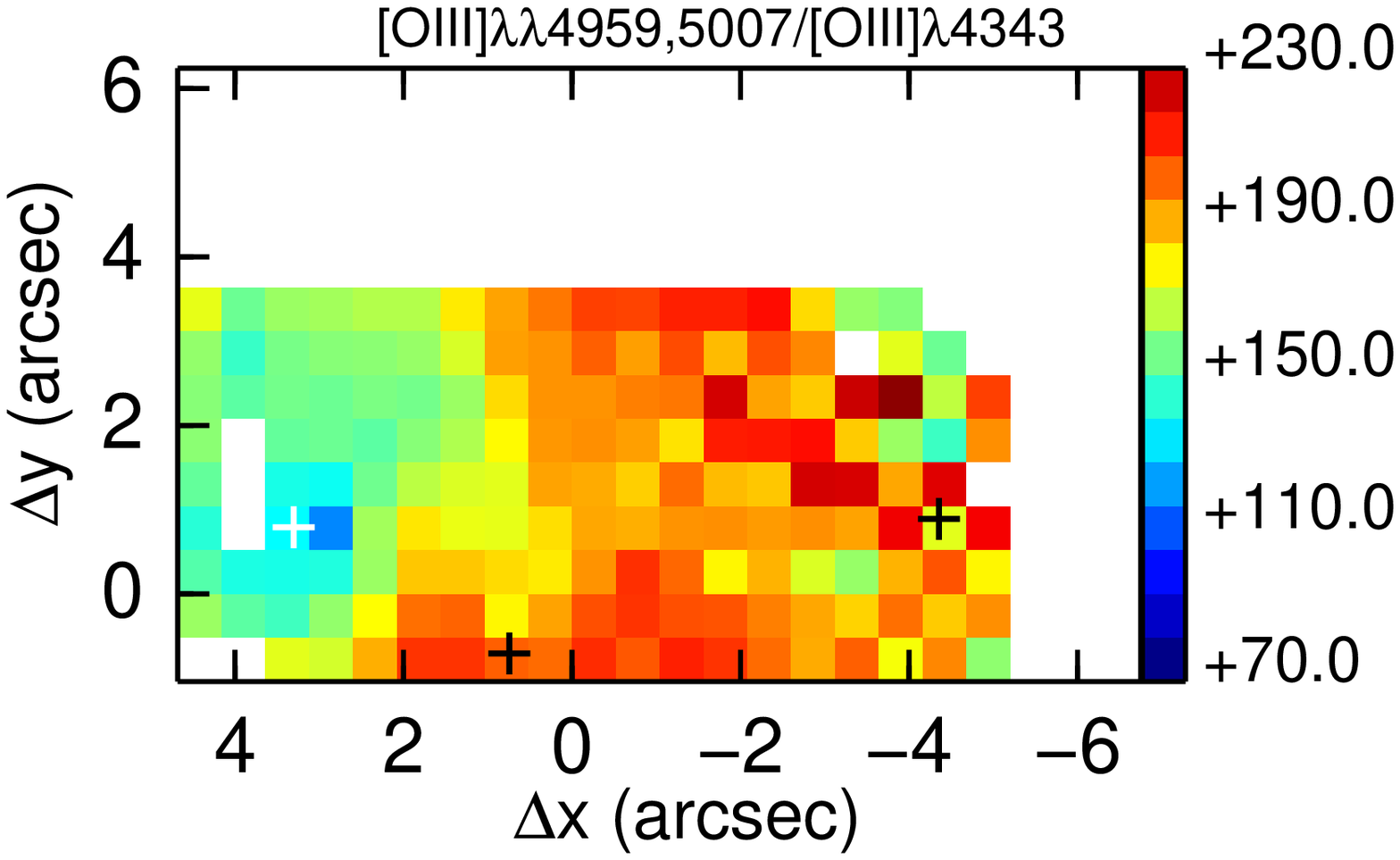}
\includegraphics[angle=0,width=0.48\textwidth,clip=, bb = 30 110 530 415]{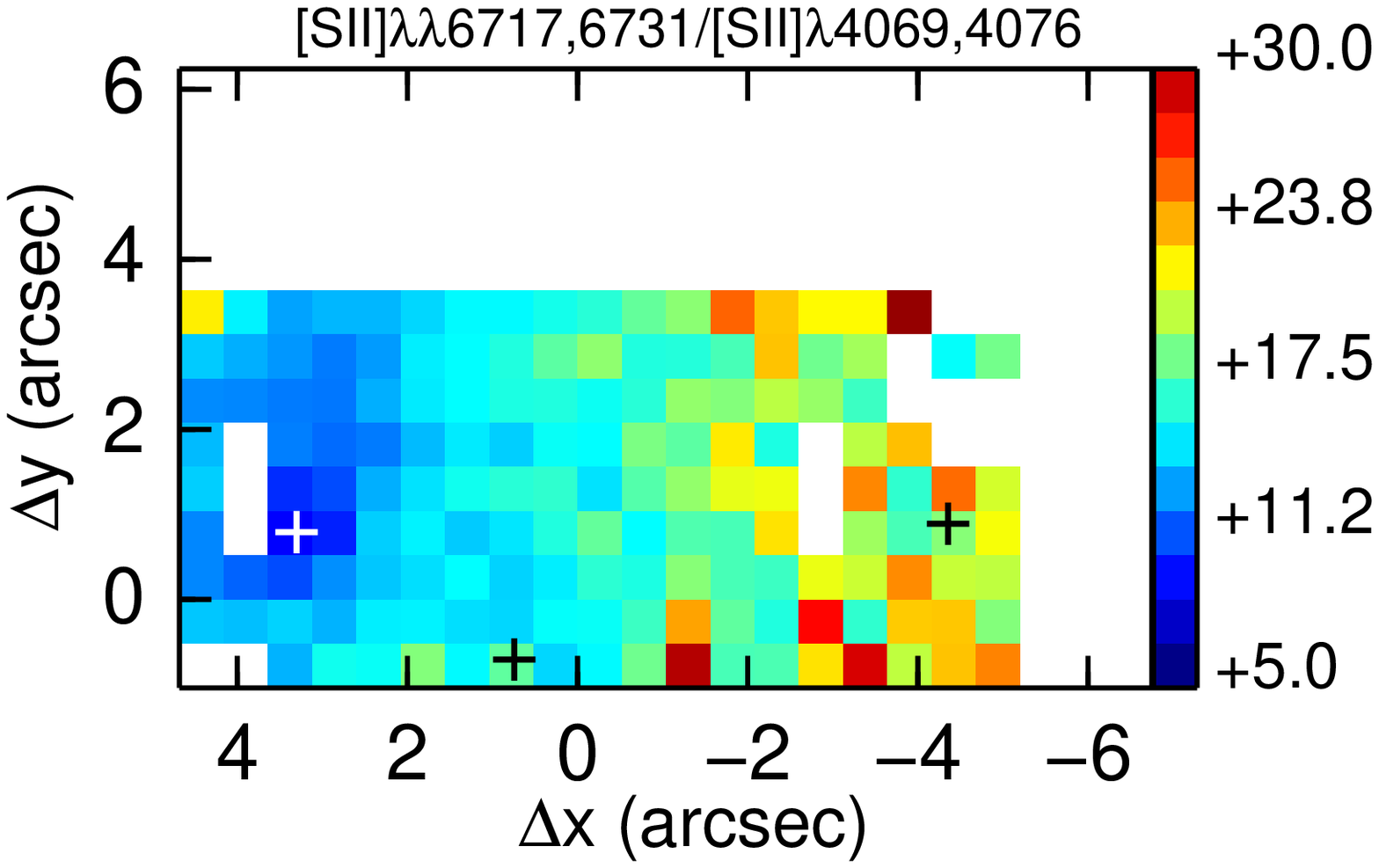}
   \caption[Maps for $T_e$ sensitive line ratios]{Maps for line ratio sensitive to $T_e$ tracing the high (\emph{up}) and the low (\emph{bottom}) ionization regime. The position of the three main peaks of continuum emission are shown as crosses for reference. 
 \label{mapo3ands2rat}}
 \end{figure}

\subsection{Mapping electron density and temperature tracers \label{mapneyte}}

In \citetalias{mon10}, we found an $n_e$(\sii) gradient declining from the peak of emission in \ha\ outwards and traced the gas at the highest densities by means of the $n_e$(\ariv). Also, our multi-component analysis of the \sii\ line profiles tentatively suggested similar densities over the whole face of the main H\,\textsc{ii} region for the broad component, while the narrow component presented somewhat lower (higher) ratios towards the NW (SE) part of the region. Given the restricted spectral range utilized in that work, the $T_e$ structure could not be derived.

Here, we present maps for two additional tracers of $n_e$: \oii$\lambda$3726/\oii$\lambda$3729 and \feiii$\lambda$4986/\feiii$\lambda$4658. In comparison to \sii$\lambda$6717/\sii$\lambda$6731, the first ratio is sensitive to a slightly larger range of densities, tracing to lower $n_e$ \citep{ost06}. Moreover, the \oii\ lines are stronger than those of \sii. Therefore, we will be able to more systematically discuss the differences in $n_e$ structure between different kinematic components, as suggested by the \sii\ line ratio.
The second ratio is sensitive to an even larger range of densities, from $\sim10^2$~cm$^{-3}$ to $\lsim10^7$~cm$^{-3}$ \citep{kee01} and has already been used with success in similar works to this one to reveal very high density locations \citep{jam09}.
Also, we will introduce for the first time maps for two tracers of $T_e$ in this galaxy: \oiii$\lambda\lambda$4959,5007/\oiii$\lambda$4343 and \sii$\lambda\lambda$6717,6731/$\lambda\lambda$4069,4076. By comparing the maps in $n_e$ and $T_e$ according to the different tracers, we will be able to see how the physical properties of the gas are structured in the different ionization layers (see Sec. \ref{discusion}). 

%
The maps for our new $n_e$ tracers are presented in Fig. \ref{mapo2rat}. For that involving the \oii\ emission lines, large values, associated to relatively large $n_e$, are found in the main \ghiir. There is a secondary maximum at the location of knot $\sharp$2 and ratios typical of low densities elsewhere. The map for the \feiii\ line ratios presents the inverse tendencies: the maximun of \oii\ line ratio corresponds to the minimum of the \feiii\ line ratio. The two overall structures are similar to the one depicted by the \sii\ line ratios \citepalias{mon10}. 

More interestingly, Fig. \ref{mapo3ands2rat} presents the maps for the line ratios tracing the $T_e$. Both of them display similar structure, with the absolute minimum at the location of the two SSCs (i.e. knot $\sharp$1), corresponding to the highest $T_e$, ratios remaining relatively low in the \ghiir and then tending to higher values in the rest of the fov. 
As with the selected apertures in Sec. \ref{fiscon}, on each individual spaxel the predicted $T_e$ from the \oiii\ line ratio differs from the one using the \sii\ lines. We will explore in Sec. \ref{discusion} how this is related to the ionization structure of the galaxy.

   \begin{figure}[th!]
   \centering
\includegraphics[angle=0,width=0.48\textwidth,clip=,bb = 30 110 530 415]{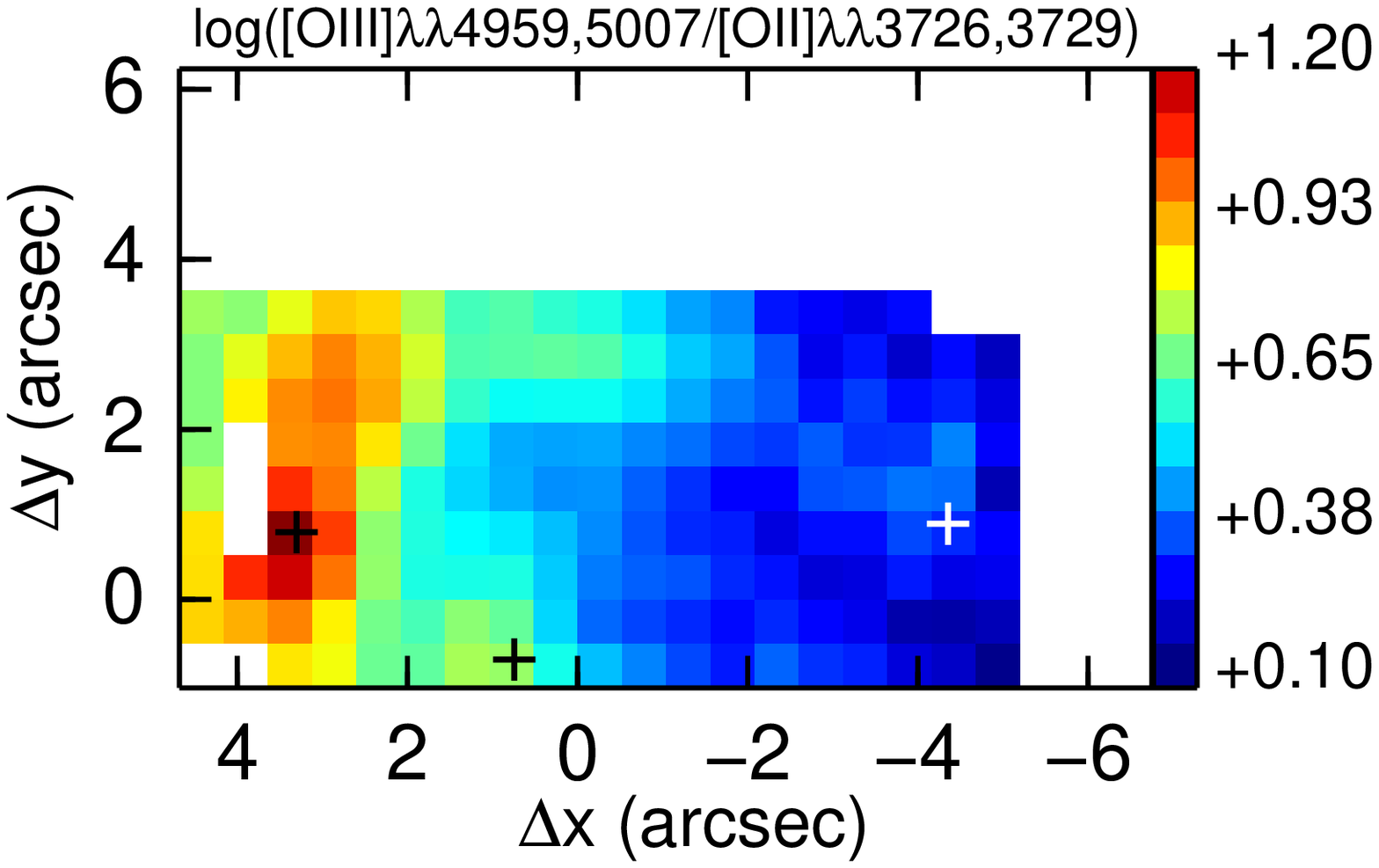}
\includegraphics[angle=0,width=0.48\textwidth,clip=,bb = 30 110 530 415]{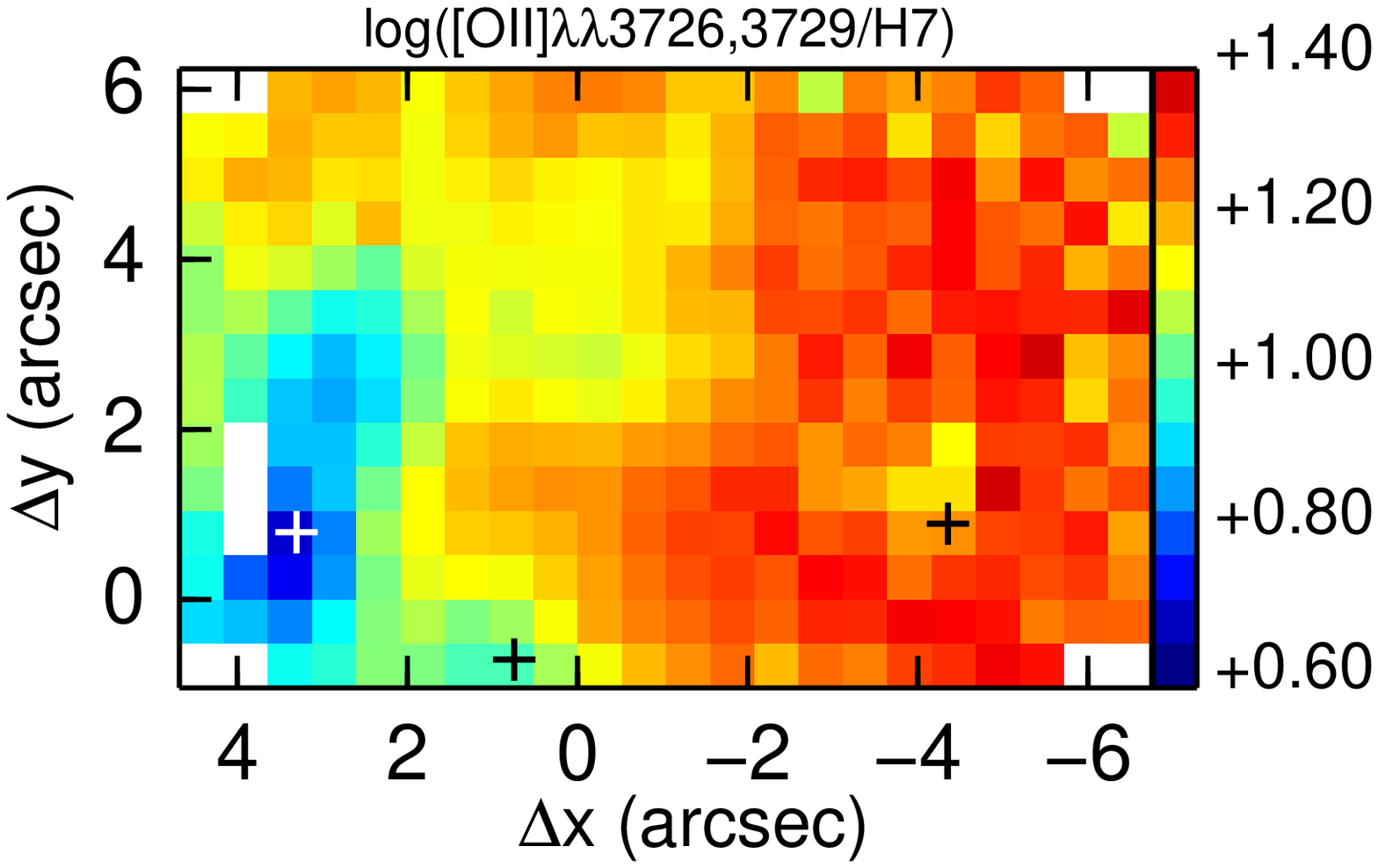}
   \caption[Maps for two tracers of the ionization degree]{Maps for two tracers of the ionization degree. \emph{Top:} \oiii/\oii.
\emph{Bottom:} \oii/H7. The position of the three main peaks of continuum emission are shown as crosses for reference. 
 \label{mapsu}}
 \end{figure}

  \begin{figure}[th!]
   \centering
\includegraphics[angle=0,width=0.40\textwidth,clip=, bb = 60 365 520 695]{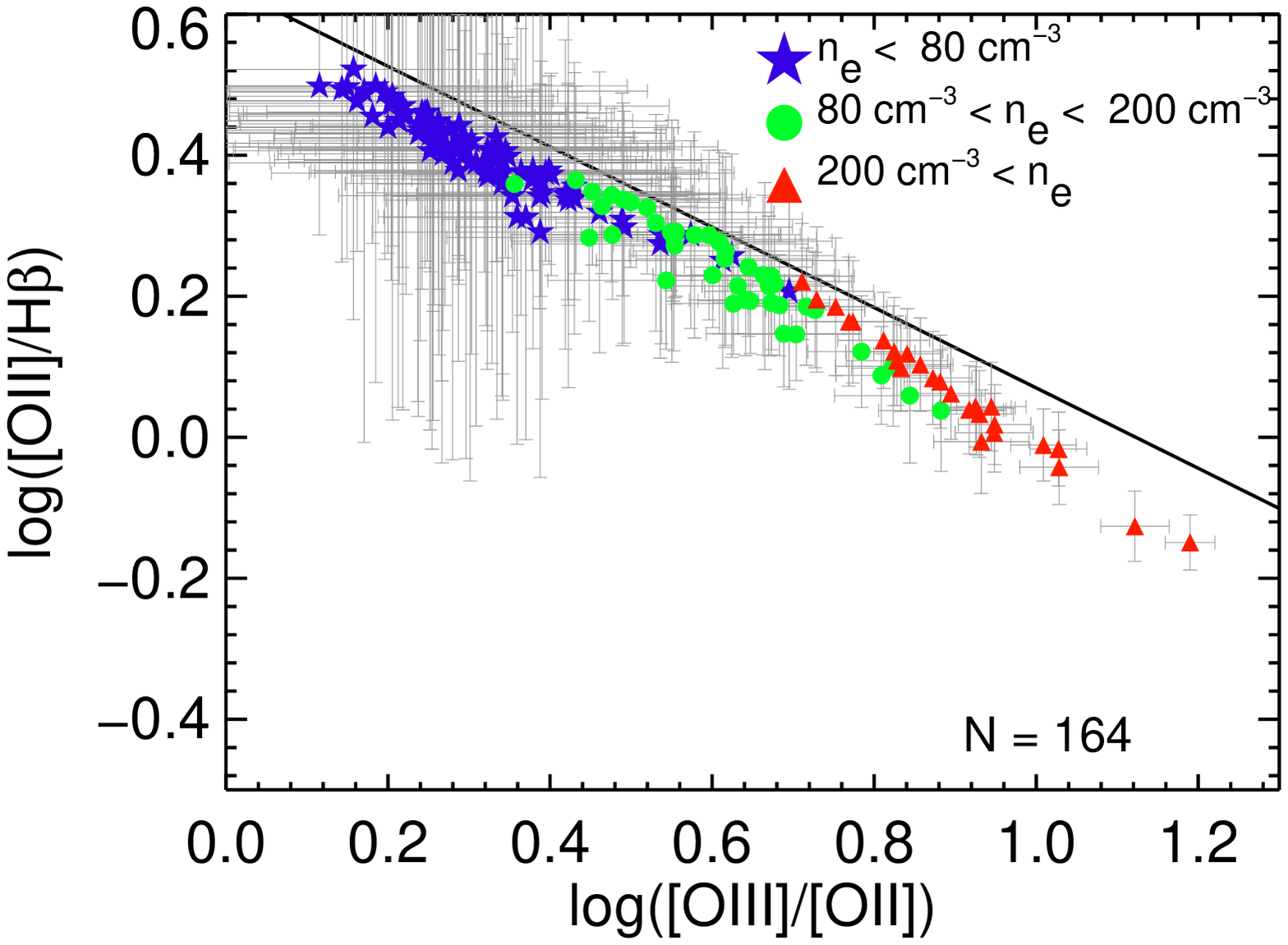}
\includegraphics[angle=0,width=0.40\textwidth,clip=, bb = 60 365 520 695]{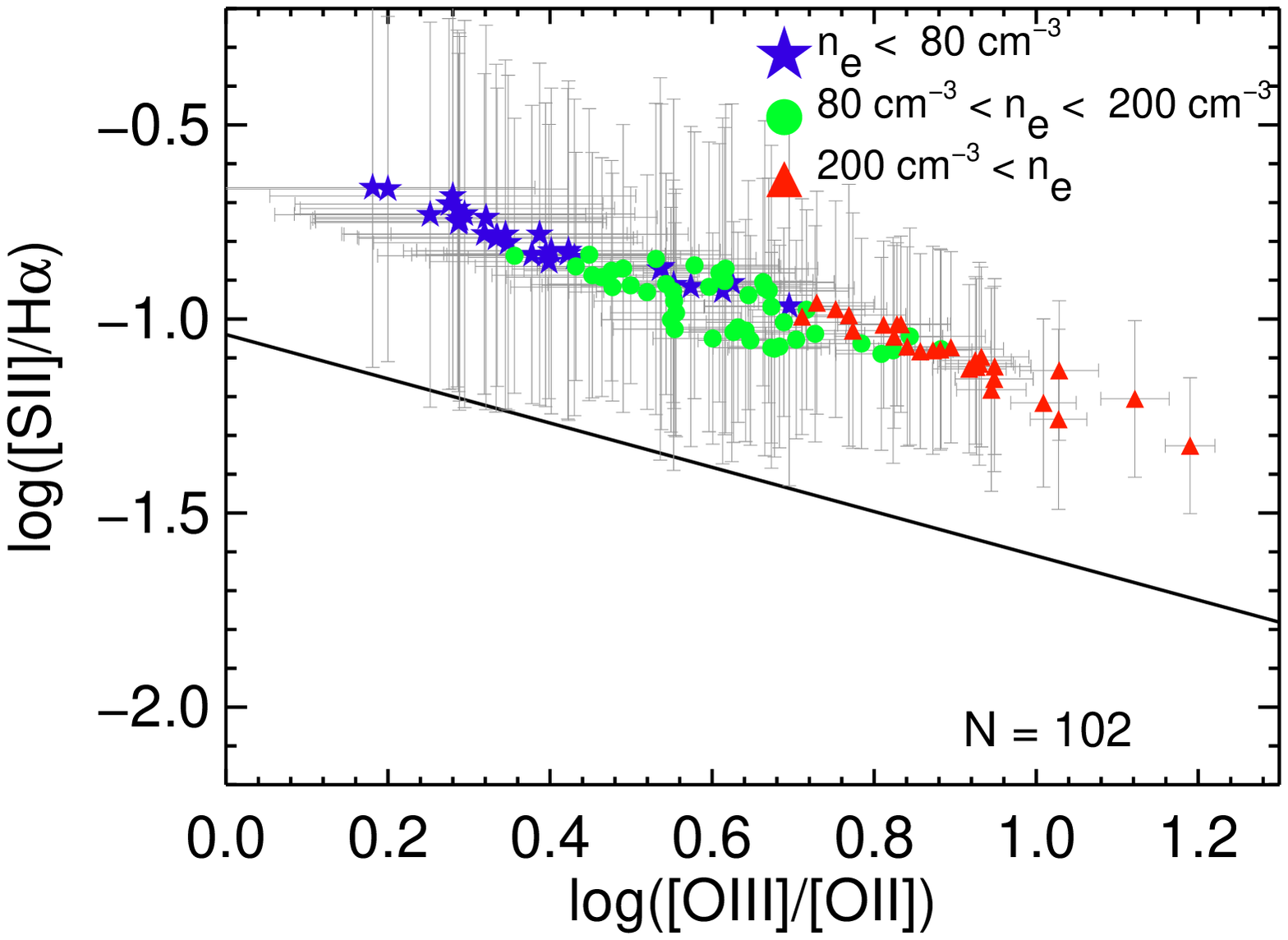}
   \caption[Diagrams involving two $U$-sensitive line ratios]{Diagrams involving two ratios sensitive to the ionization degree. \emph{Top:} \oiii$\lambda\lambda$4959,5007/\oii$\lambda\lambda$3726,3729 vs. \oii$\lambda\lambda$3726,3729/H$\beta$. \emph{Bottom:} \sii$\lambda\lambda$6717,6731/H$\alpha$ vs. \oiii$\lambda\lambda$4959,5007/\oii$\lambda\lambda$3726,3729).
The locus of equal estimated ionization parameter according to the relations proposed by \citet{dia00} and assuming $Z=0.3\,Z_\odot$ is indicated with a black line. 
Only those spaxels with estimated uncertainties for the involved ratios smaller than 0.5~dex have been taken into account.
Spaxels were grouped in three bins according their $n_e$(\oii) as indicated in the upper right-hand corner of the individual diagrams. The total number of considered data points is indicated in the lower right-hand corner. 
 \label{compau}}
 \end{figure}

\subsection{Mapping tracers of the local ionization degree\label{localioni}}

The relation between the \nha\ and \sha\ line ratios in galaxies is not monotonic, even though the $S^+$ and $N^+$ ionization potentials and critical densities are in a similar range.
Instead, different ionization mechanisms \citep[shocks, star formation, AGNs, e.g. ][]{mon06,mon10b} and/or physical and chemical conditions of the gas (metallicity, relative abundances and degree of ionization) trace different loci in the \nha\ vs. \sha\ diagram. In \citetalias{mon10}, we attributed the different structure in the \nha\ and \sha\ maps to an excess in nitrogen. This was supported by evidence of star-formation as the main ionization mechanism and previous estimations of the ionization parameter at specific locations based on long-slit measurements \citep{kob97}. However, certain combinations of specific ionization structures and metallicity gradients could reproduce a similar locus in the \nha\ vs. \sha\ diagram. This possibility can be ruled out with the present data.
Here, we present the ionization structure in \object{NGC~5253} as traced by different line ratios, while the next section will contain maps for the abundances of different heavy elements (oxygen, nitrogen, neon and argon). 

The degree of ionization can be traced by means of ratios of lines of the same element tracing two different ionization states. The upper panel of Fig. \ref{mapsu} contains a map for one of these ratios: \oiiioii. Note that in order to minimize uncertainties associated to aperture matching, absolute flux calibration and extinction, the \oiii\ and \oii\ lines were measured with respect to \hb\ and H7. Then, we assumend the 
theoretical Balmer line intensities obtained from \citet{sto95} for Case B, $T_e= 10^4$ K and $n_e=$ 100 cm$^{−3}$. The map shows that the ionization structure reproduces the morphology observed for the ionized gas. That is: i) line ratios tracing the highest ionization degree are associated to knot $\sharp$1; ii) relatively high ionization is found in the main \ghiir; iii) a secondary peak of high ionization is found around knot $\sharp$2; iv) low ionization degree is found in the rest of the fov, where the diffuse component of the ionized gas becomes more relevant. 

Also, if the metallicity is known, line ratios like  \oii$\lambda\lambda$3726,3729/H7 or  \sha\ can also be used to trace the ionization degree. The map \oii$\lambda\lambda$3726,3729/H7 is presented in the lower panel of Fig. \ref{mapsu} while that of \sha\ was included in \citetalias{mon10}. The similar structure in all the three maps suggests a lack of metallicity gradient in the galaxy as will be shown in the next section.

In order to explore whether these three tracers predict consistent ionization degree for a given position, we show in Fig. \ref{compau}, the relation between the different ratios for each individual spaxel, using \oiiioii\ as reference. Also, we overplotted the locus of equal estimated ionization parameter according to the relations proposed by \citet{dia00} and assuming $Z=0.3\, Z_\odot$ with a black line. For a given location, \oiiioii\ and \oii$\lambda\lambda$3726,2729/\hb\ predict a similar degree of ionization while the predictions of \sha\ would correspond to smaller ionization parameters. A similar result was found in a detailed analysis of NGC~588, a \ghiir\ in M~33 \citep{mon11}. Both, the spatial variations of the different line ratios (Fig. \ref{mapsu}) and the observed excess in the \sha\ ratio when compared with photoionization models (Fig. \ref{compau}) can be jointly explained as a 3D view of the ionization structure of the galaxy. Specifically, for a given spaxel (i.e. a given line of sight), the lower ionization species (e.g. $S^+$), delineate the more extended diffuse component while $O^{++}$ will be confined to different high ionization zones closer to the ionizing sources. Alternatively, the difference between the ratios found here and those predicted by the D\'{\i}az et al.  models can be attributed to differences between the relative $S/O$ abundances in \object{NGC~5253} \citep[$\log(S/O)\sim-1.47$,][]{kob97,sid09} and those assumed in the models \citep[$\log(S/O)=-1.71$,][]{gre89}. A detailed modelling of the ionization structure of the galaxy - out of the scope of this work - could help to disentangle these two possibilities.

\begin{figure*}[th!]
   \centering
\includegraphics[angle=0,width=0.48\textwidth,clip=, bb = 30 110 530 415]{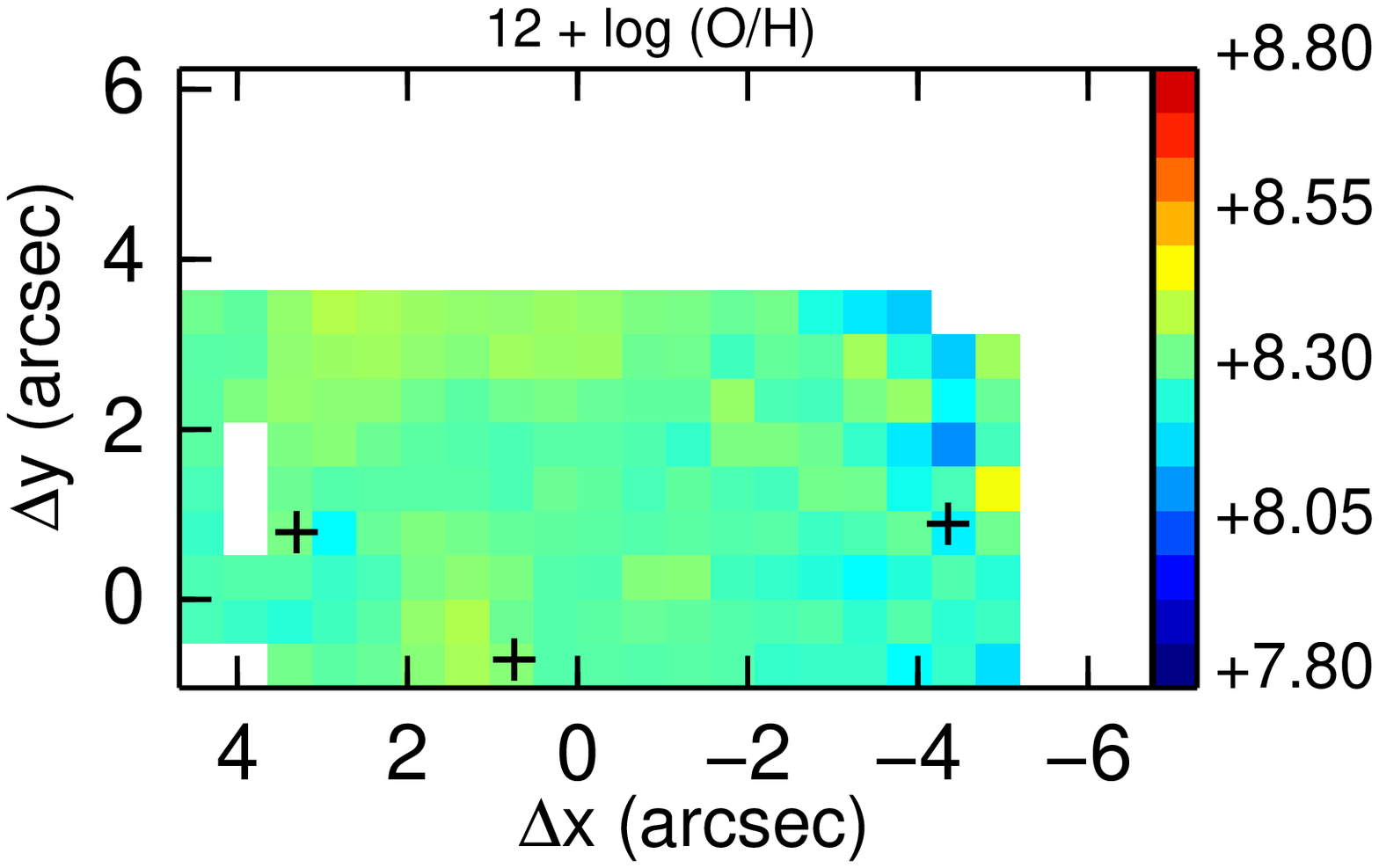}
\includegraphics[angle=0,width=0.48\textwidth,clip=, bb = 30 110 530 415]{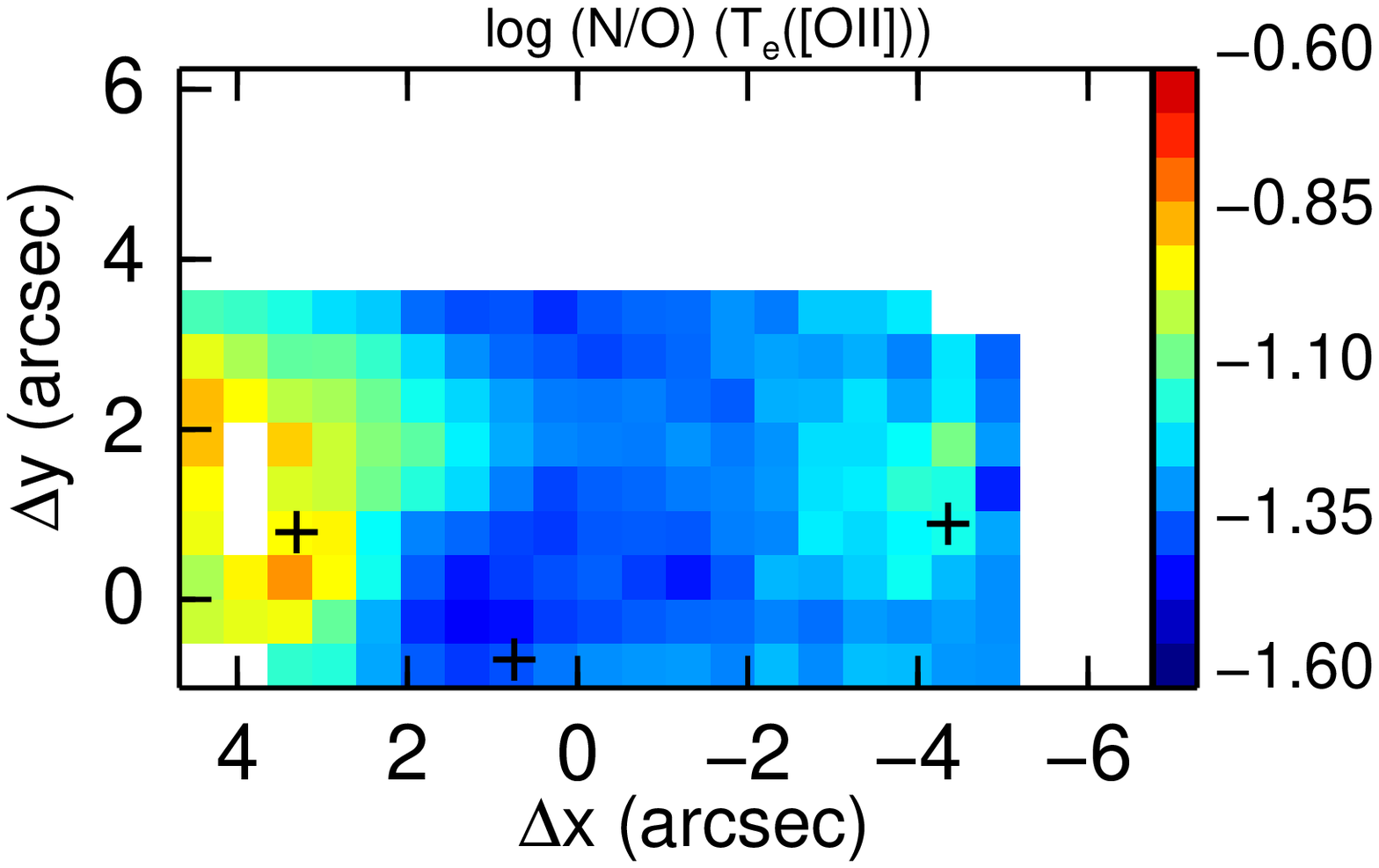}\\
\includegraphics[angle=0,width=0.48\textwidth,clip=, bb = 30 110 530 415]{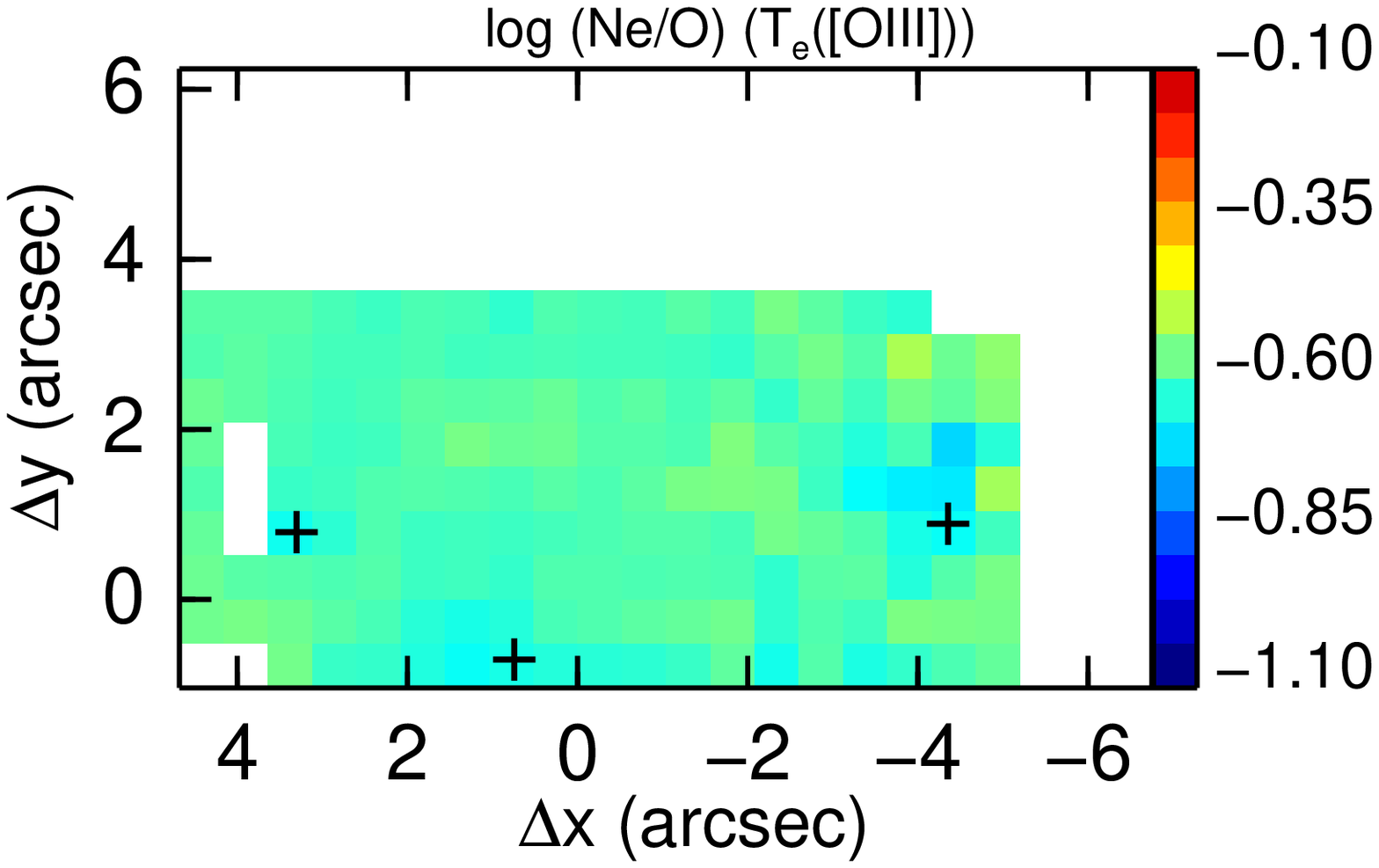}
 \includegraphics[angle=0,width=0.48\textwidth,clip=, bb = 30 110 530 415]{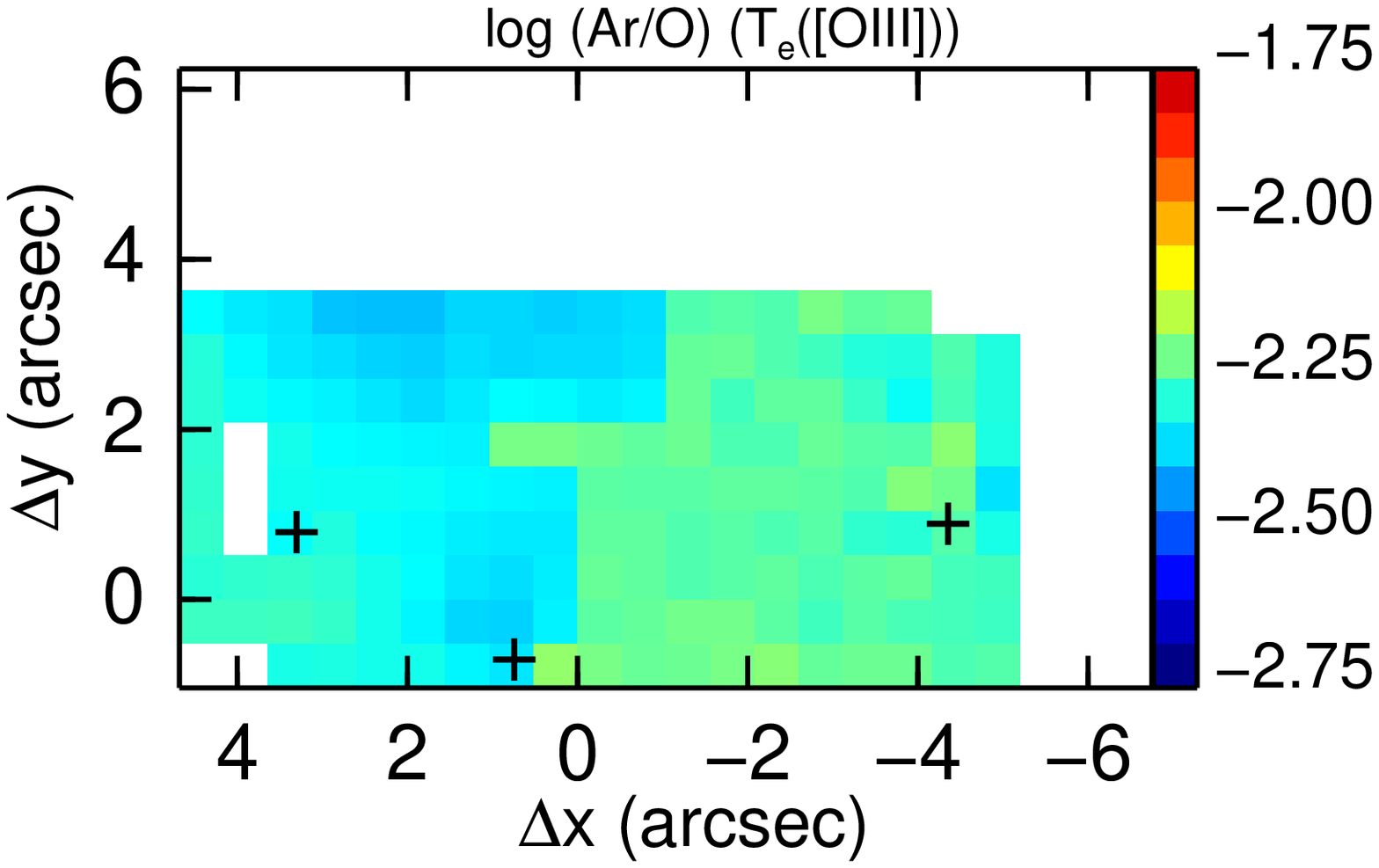}
  \caption[Metallicity and relative abundances maps]{Metallicity and relative abundances maps. From top to bottom and left to right: Oxygen abundance, and nitrogen, neon and argon relative abundances with respect to oxygen. In order to make easier the comparison between the different maps, all of them cover a range of 1~dex.
 \label{mapabun}}
 \end{figure*}

\subsection{Mapping abundances of heavy elements \label{secabun}}

Following the same methodology as in Sec. \ref{abunaper}, we derived maps for the abundances of oxygen, nitrogen, neon and argon. These are presented in Fig. \ref{mapabun} and show, as expected, a homogeneous distribution within the uncertainties in all the elements but nitrogen.
Note that the spaxel-to-spaxel variations in these maps are not dominated by the quality of the data (e.g. S/N ratio), but are inherent to the adopted $T_e$ (and to a much lesser extent $n_e$) as well as the utilized methodology. The most obvious example would be the map showing the distribution of the argon abundance, where a different methodology was adopted to estimate the ICF depending on whether the \ariv$\lambda$4740 emission line was or was not detected \citepalias[see Fig. 6 in][to locate those spaxels with detection]{mon10}. Nevertheless, abundances derived in these two groups differ only by $\sim$0.1~dex, well within the expected uncertainties for this element. Also, $12+\log(O/H)$ marginally anti-correlates with $T_e$(\oiii) in the spaxels with lowest surface brightness.
The mean ($\pm$standard deviation) metallicity in our fov is $12+\log(O/H)=8.26$ $(\pm0.04)$ while the relative abundances $\log(Ne/O)$ and $\log(Ar/O)$ are  $-0.64 (\pm0.03)$ and $-2.32 (\pm0.05)$.
Supporting this homogeneity, in all three cases -- oxygen, neon and argon -- the distribution of the values measured over the fov can be well reproduced by a Gaussian with $\sigma\sim0.03-0.07$~dex and all the values measured in the individual spaxels are consistent with the mean value at the 3$\sigma$ level. 

The abundance in nitrogen differs from this pattern. As reported in \citetalias{mon10}, there is a roughly elliptical area of $\sim80$~pc$\times35$~pc, associated to the main H\,\textsc{ii} region and centered on the two massive SSCs, that presents an enhancement of $N/O$. 
More interestingly, a putative secondary excess in nitrogen not reported so far appears in the vicinity of (but not centered on) knot $\sharp$3, at the area of our fov with the lowest surface brightness. 

The existence of a second area with slight $N/O$ enhancement is supported by the histogram of the values measured along the FLAMES fov (not shown). 
Contrary to the case for $O$, $Ne$ and $Ar$, three Gaussians centered at $\log(N/O)=-1.32$, $-1.17$ and $-0.95$ and with widths $\sigma= 0.05,0.07,0.03$ are needed to reproduce the full distribution. Most of the spaxels are associated to the first Gaussian and trace the typical $N/O$ abundance for \object{NGC~5253}. The second largest group, with $\log(N/O)\sim-0.95$, are associated to the area with N-enhancement already reported for the main \ghiir.
Using the criterion that spaxels belonging to two Gaussian trace different abundances if their centers are separated by more than 3$\sigma$, we can conclude that the third group of spaxels, which are associated to the area in the vicinity of knot $\sharp$3, traces a zone with slightly larger, but clearly distinct, $N/O$ value than the typical one for the galaxy. 
Given that this area is located at the spaxels with the lowest surface brightness, and to reject the possibility of any systematic effect due to a poor S/N ratio in faint lines like \oiii$\lambda$4363, we extracted a spectrum by summing up the flux in a rectangular area of 5$\times$6 spaxels at this location. The measured relative N-abundance, $\log(N/O)=-1.15$, is in agreement with the result found on a spaxel-by-spaxel basis.

\begin{figure}[th!]
   \centering
\includegraphics[angle=0,width=0.48\textwidth,clip=, bb = 30 110 530 415]{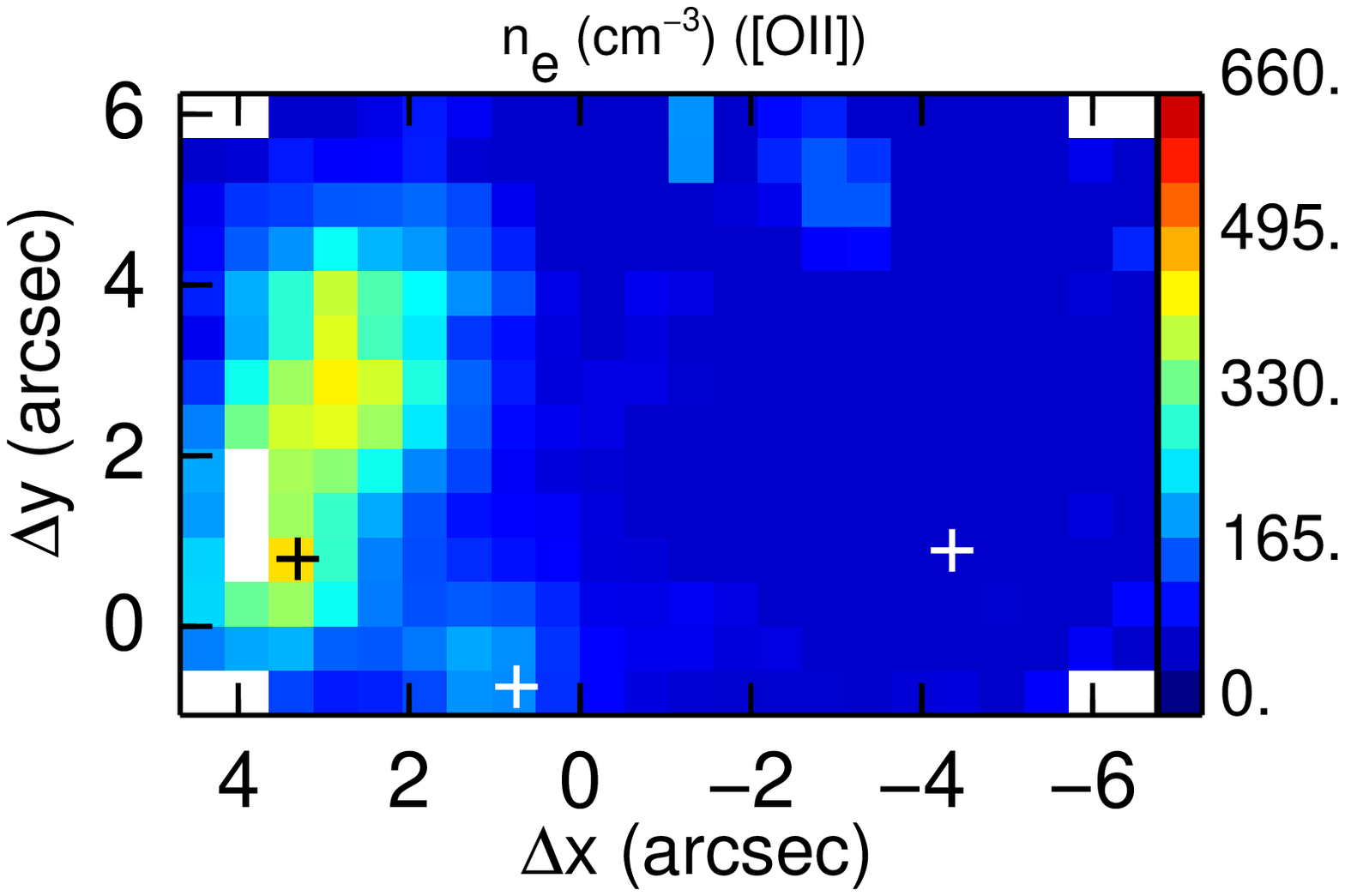}
\includegraphics[angle=0,width=0.48\textwidth,clip=, bb = 30 110 530 415]{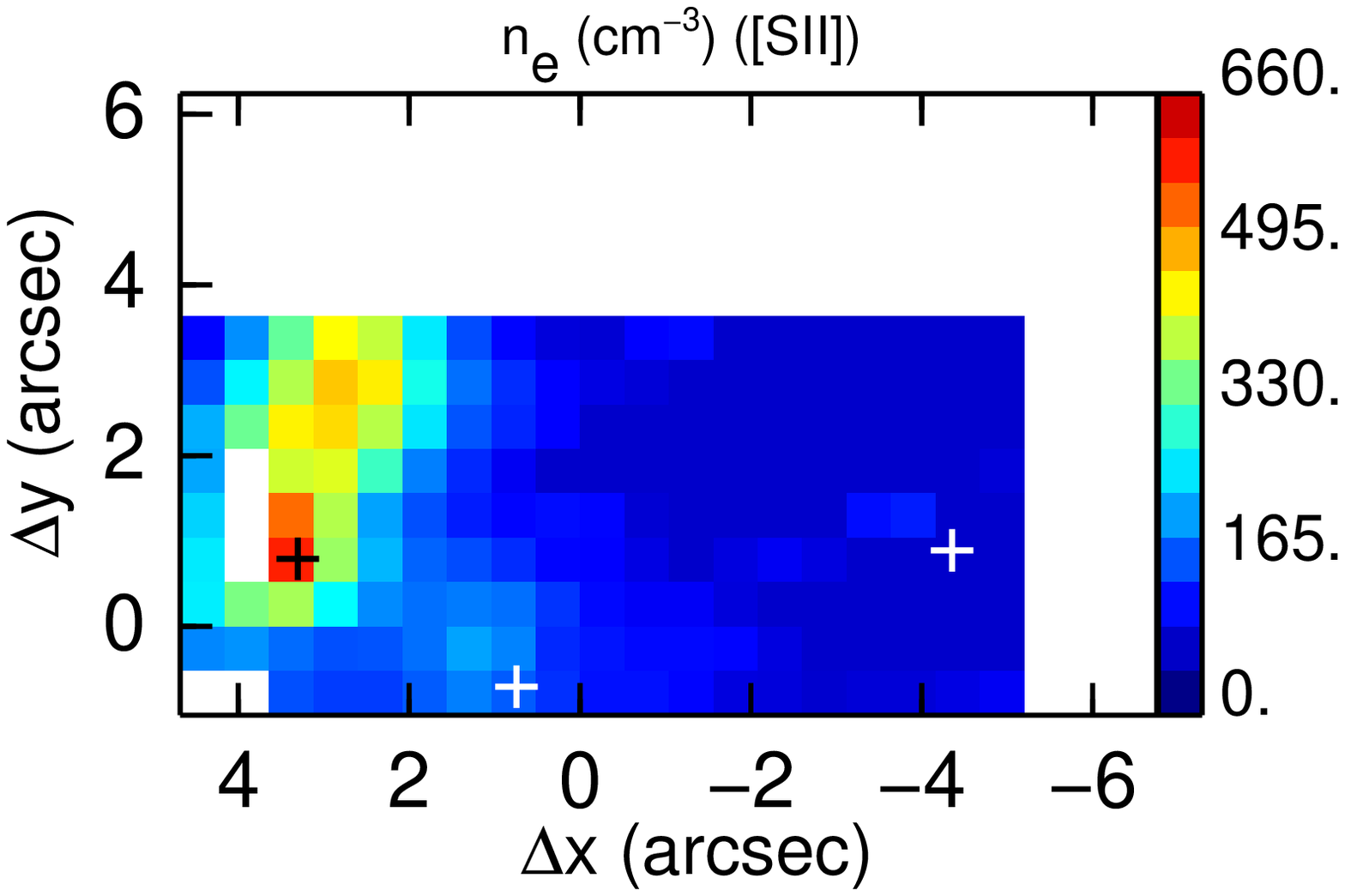}
\includegraphics[angle=0,width=0.48\textwidth,clip=, bb = 30 110 530 415]{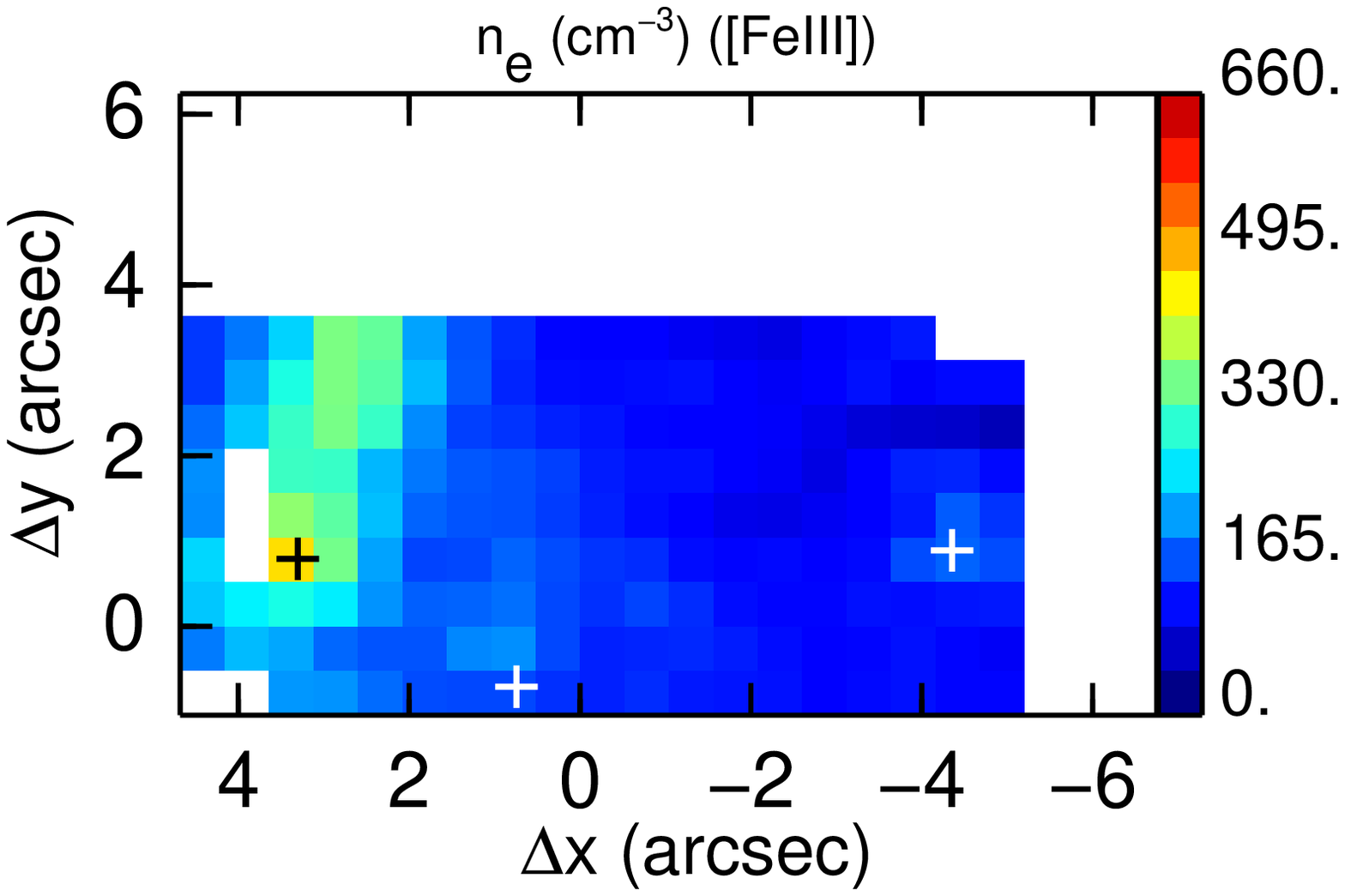}
   \caption[Electron density maps from different tracers]{Electron density maps from different tracers. From top to bottom: \oii,  \sii, and \feiii. The position of the three main peaks of continuum emission are shown as crosses for reference. 
 \label{mapne}}
 \end{figure}

   \begin{figure}[th!]
   \centering
\includegraphics[angle=0,width=0.48\textwidth,clip=, bb = 30 110 530 415]{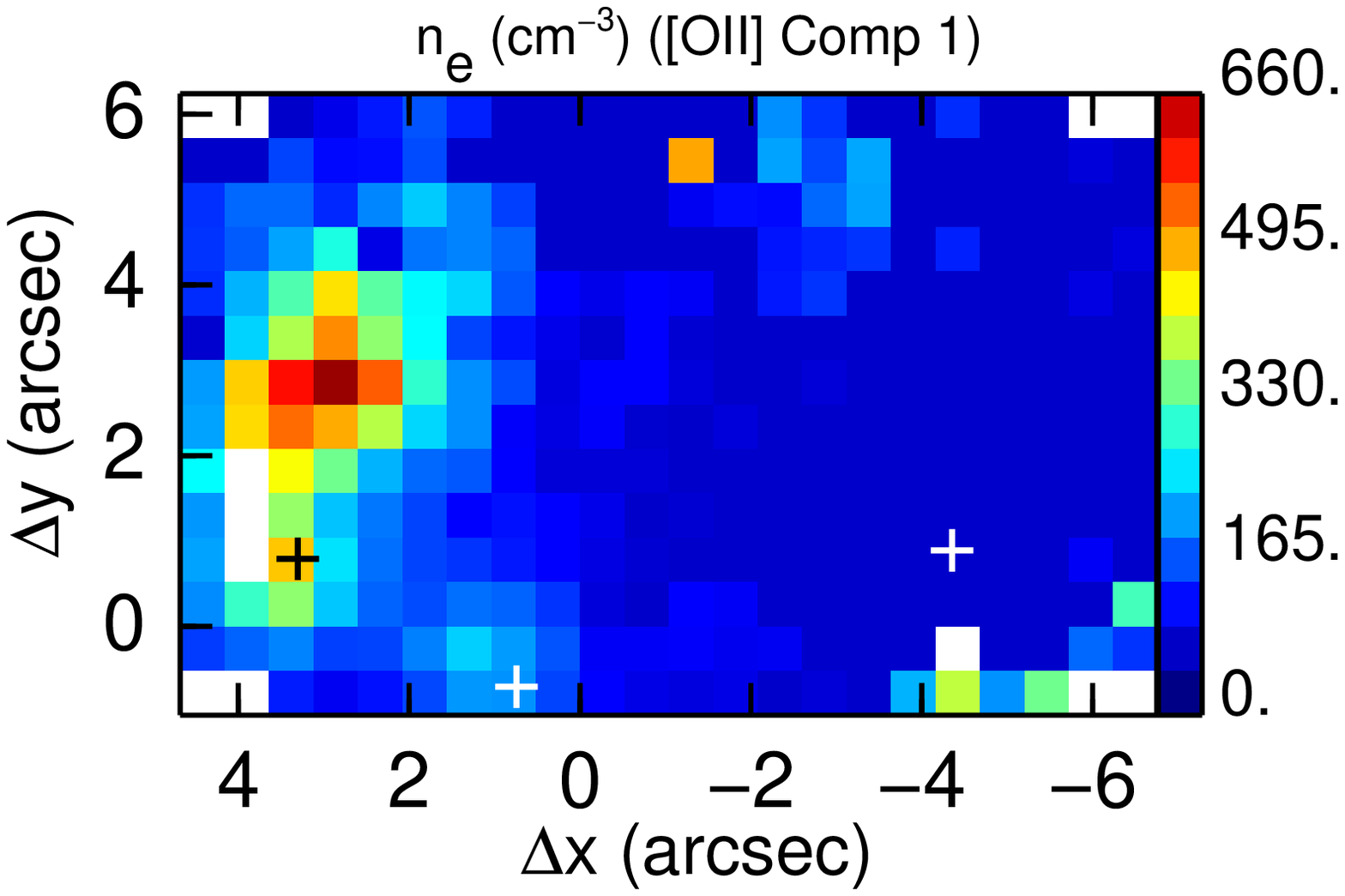}
\includegraphics[angle=0,width=0.48\textwidth,clip=, bb = 30 110 530 415]{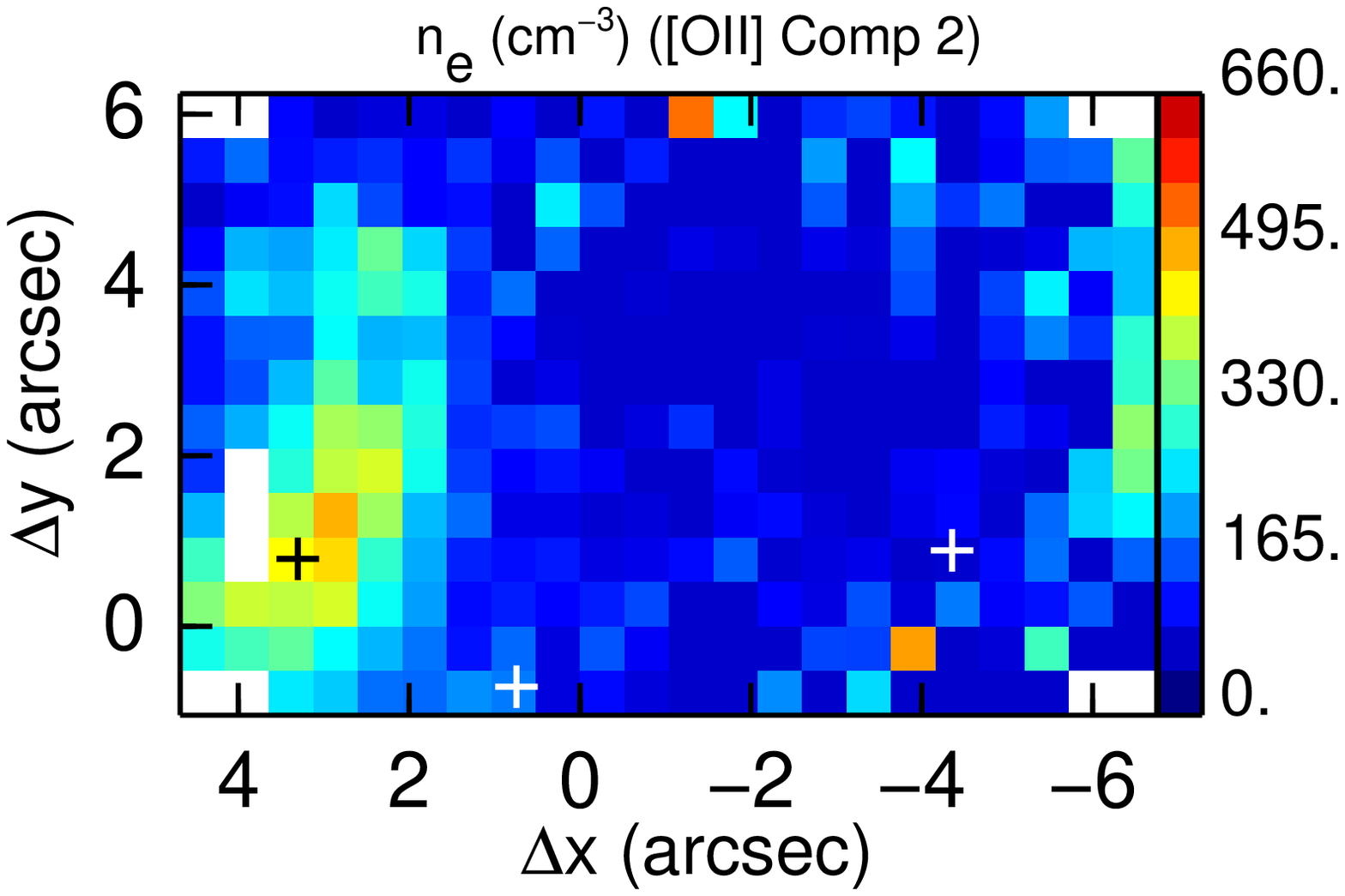}
   \caption[Electron density maps for the different kinematical components $\[\oii\]$]{Electron density maps for the different components using the results for \oii. \emph{Top:} Narrow component; \emph{Bottom:} Broad component. The position of the three main peaks of continuum emission are shown as crosses for reference. 
 \label{mapnemulti}}
 \end{figure}

   \begin{figure}[th!]
   \centering
\includegraphics[angle=0,width=0.48\textwidth,clip=, bb = 30 110 530 415]{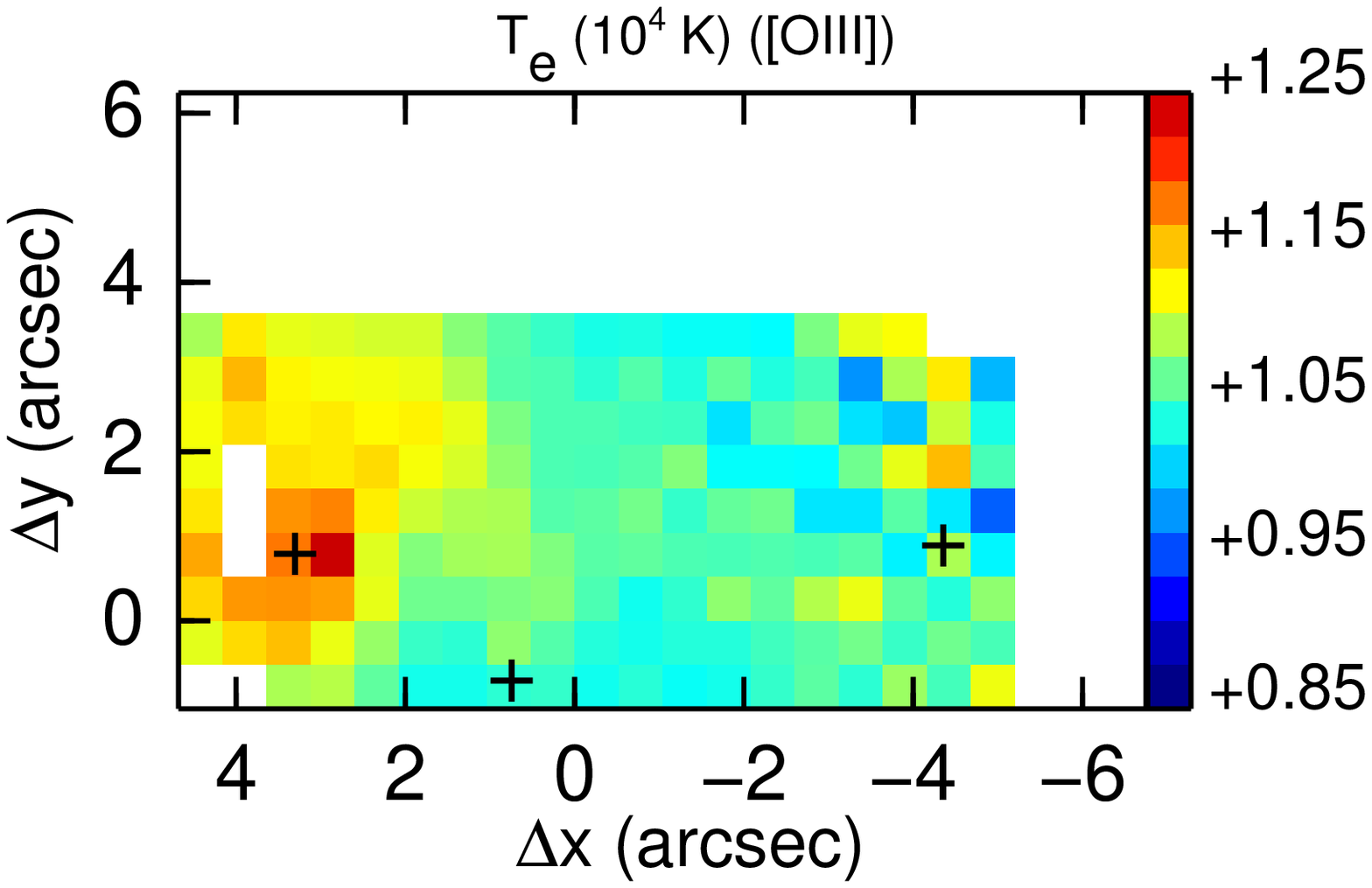}
\includegraphics[angle=0,width=0.48\textwidth,clip=, bb = 30 110 530 415]{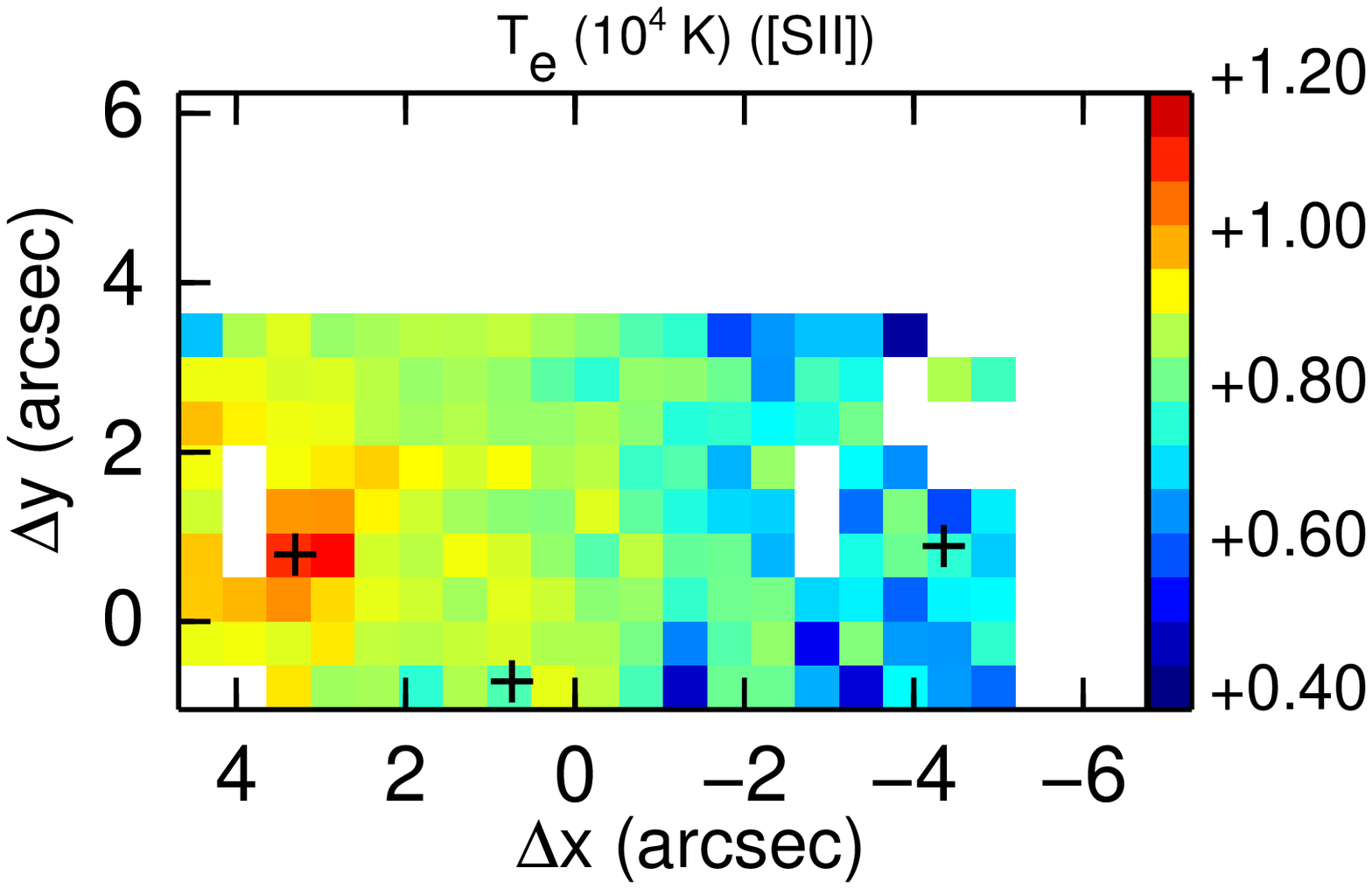}
   \caption[Electron temperature maps from  different tracers]{Electron temperature maps from different tracers. \emph{Top:} \oiii; \emph{Bottom:} \sii. The position of the three main peaks of continuum emission are shown as crosses for reference. 
 \label{mapte}}
 \end{figure}

   \begin{figure}[th!]
   \centering
\includegraphics[angle=0,width=0.48\textwidth]{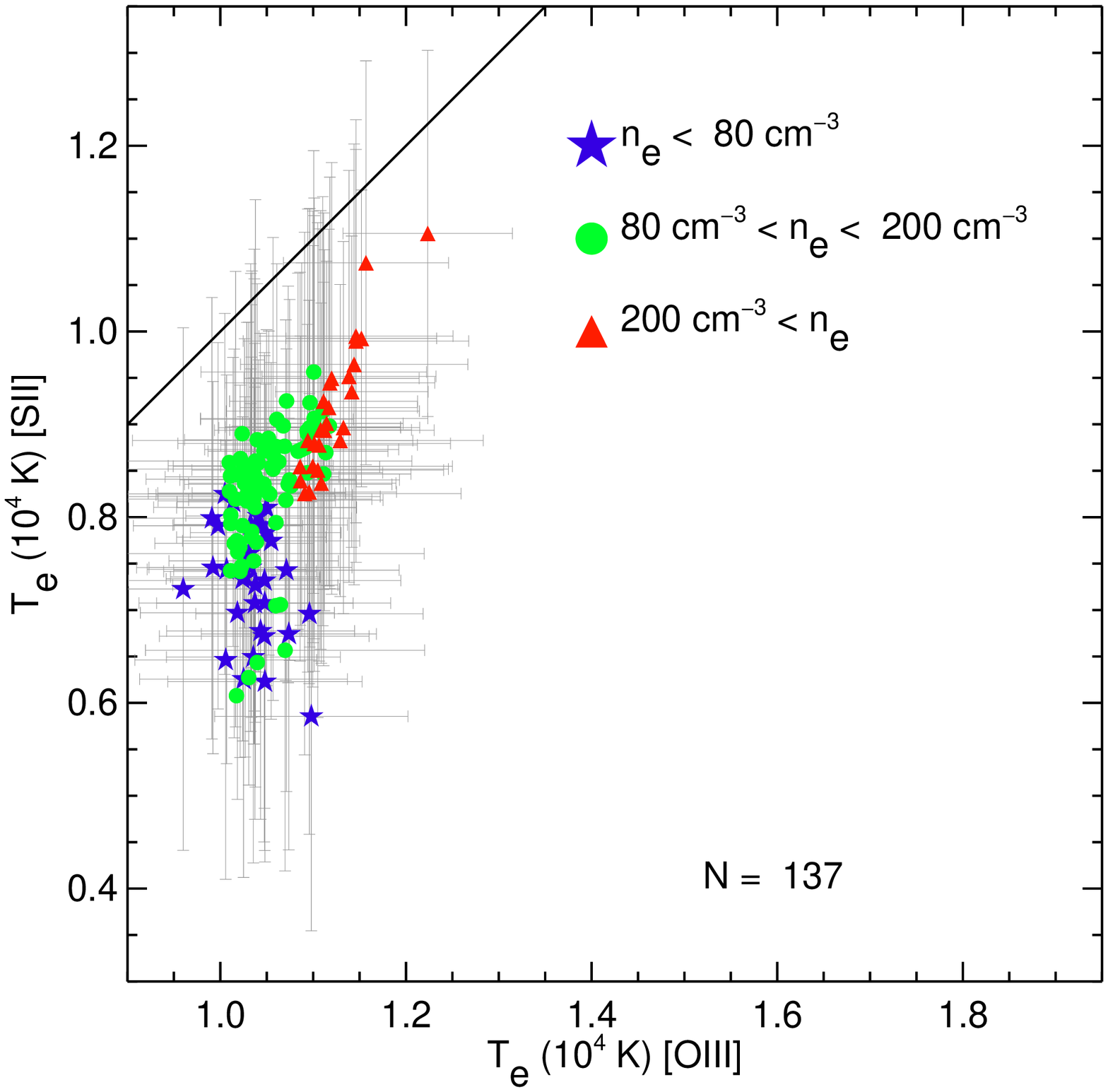}
    \caption[Comparison between the predicted $T_e$(\oiii) and $T_e$(\sii).]{Comparison between the predicted $T_e$(\oiii) and $T_e$(\sii). Spaxels were grouped in the same three bins in $n_e$(\oii) as in Fig. \ref{compau} as it is indicated in the upper right-hand corner of the diagram.
\label{compate}}
 \end{figure}

\section{Discussion \label{discusion}}

\subsection{Electron density structure \label{densistruc}}

As we discussed in Sec. \ref{localioni}, for a given spaxel, different ions are associated to different layers along the line of sight. 
We have presented maps for tracers of electron density based on \sii, \feiii\ and \oii\ emission line ratios \citepalias[][ and Sec. \ref{mapneyte}]{mon10}.
%
The corresponding density maps were derived using the same methodology as in Sec. \ref{fiscon} and appear in Fig. \ref{mapne}. Values consistent with being below the low-density limit are found in most of the fov, independently of the utilized tracer, while knot $\sharp$2 presents values somewhat larger ($\sim$190~cm$^{-3}$). The richest density structure is found in the main \ghiir. Here, all three tracers depict an $n_e$ structure with two peaks: the first one centered at knot $\sharp$1 while the second one at $\sim$2\farcs6 ($\sim$50~pc) towards the northwest. However the values of the peaks vary depending on the utilized tracer. The largest densities if Fig. \ref{mapne} are traced by \sii, while \oii\ and \feiii\ predict densities about 10-25\% smaller.
Given their ionization potentials, (23.3, 35.1, 30.7 eV for \sii, \oii\ and \feiii, respectively), in the absence of extinction, $n_e$(\oii)  - for example - traces the density in a layer closer to the ionizing source than $n_e$(\sii) \citep[see e.g.][]{gar92}. This can be reversed in conditions of heavy extinction. Even if the intrinsic structure would be the same, lines involved in the determination of $n_e$(\oii) are bluer (and therefore more sensitive to extinction) than those associated to $n_e$(\sii) which therefore can probe deeper in the nebula. Since this \ghiir\ suffers from relatively high extinction \citepalias[see e.g. Fig 3. in][]{mon10} the differences found between the three density maps are consistent with an onion-like structure where inner layers are denser than the outer ones. This is supported by the relative densities found between the two peaks.
Also it is consistent with the fact that the \feiii\ ratio does not trace very high densities as the \ariv\ ratio did \citepalias{mon10}. With an ionization potential of 59.8~eV, \ariv\ would sample the densest innermost layer of this onion-like structure.

The one Gaussian fitting approach utilized up to now is useful to have a picture of the overall density structure in the galaxy. However, it is known that emission lines in NGC~5253 present complex profiles with  asymmetries tracing different kinematics components.
For the \ghiir, we tentatively measured the electron density for what we called the "broad" and "narrow" components on a spaxel-by-spaxel basis  in \citetalias{mon10} using the \sii\ lines. Results were affected by large uncertainties, mainly due to the deblending procedure but also to the S/N ratio of the spectra. They suggested large and uniform densities for the "broad" component ($n_e\sim470$~cm$^{-3}$), while the "narrow" component presented a decrease of density from the northwest to the southeast. The stronger \oii\ lines offer the possibility of discussing this difference with much reduced uncertainties. Electron density maps from our multi-component analysis are presented in Fig. \ref{mapnemulti}. The structure depicted in the map for broad component (bottom panel), with larger densities ($\sim500$~cm$^{-3}$) close to knot $\sharp$1 and decreasing outwards, is consistent with the general picture sketched above (i.e. larger densities closer to the ionizing source). The narrow component (upper panel) present a different structure, with the highest density ($\lsim600$~cm$^{-3}$) at the secondary peak of the \ghiir\ at $\sim2\farcs6$ from knot $\sharp$1. This can be understood within the proposed scenario in \citetalias{mon10} (see their Fig. 21, where different elements associated with the area of the \ghiir\ are shown). There, this narrow component was associated to a shell of previously existing quiescent gas that has been reached by the ionization front. Densities in the shell would be high, if this pre-existing gas had been piled-up by the outflow associated to the broad component.
 
\subsection{Electron temperature structure}

As it happened with $n_e$, a comparison of $T_e$ maps derived from lines associated to different ions is useful to asses the $T_e$ structure in 3D of \object{NGC~5253}. Derived maps for $T_e$(\oiii) and $T_e$(\sii) are presented in Fig. \ref{mapte}. At present, maps of $T_e$ based on any tracer are still scarce and most of the times focused on Galactic objects \citep[e.g.][]{nun12,tsa08}. Indeed, to our knowledge this is the first time that a map for $T_e$(\sii) is provided and one of the few existing examples of $T_e$(\oiii) maps in extragalactic astronomy. 
As we pointed out in Sec. \ref{mapneyte}, the overall structure is the same for $T_e$(\oiii) and $T_e$(\sii) maps. However, $T_e$(\sii) is smaller than $T_e$(\oiii), with typical $T_e$(\sii)/$T_e$(\oiii) ranging from $\sim$0.8 at the \ghiir\ to $\sim$0.6 in the areas of low surface brightness.

This relation between both temperatures is shown on a spaxel-by-spaxel basis in Fig. \ref{compate}. Only those spaxels where the estimated uncertainties for both temperatures are smaller than 40\% were considered.
Both, the different values for $T_e$(\oiii) and $T_e$(\sii) and the larger differences at low surface brightness, can be understood within the frame of a change in the relative contribution along the line of sight of the warmer gas associated to the \ghiir\, and a colder and more diffuse gas component that extends further away.
The 2D information on the plane of the sky (i.e. the maps in Fig. \ref{mapte}) fits also in a satisfactory manner with this interpretation. 

Summarizing, both the available information on the plane of the sky (Fig. \ref{mapte}) and along the line of sight (Fig. \ref{compate}) are consistent with a $T_e$ structure in 3D with higher temperatures close to the ionizing source surrounded by a more diffuse component of ionized gas at lower temperatures.

\begin{figure}[th!]
   \centering
\includegraphics[angle=0,width=0.48\textwidth,clip=, bb = 30 110 530 415]{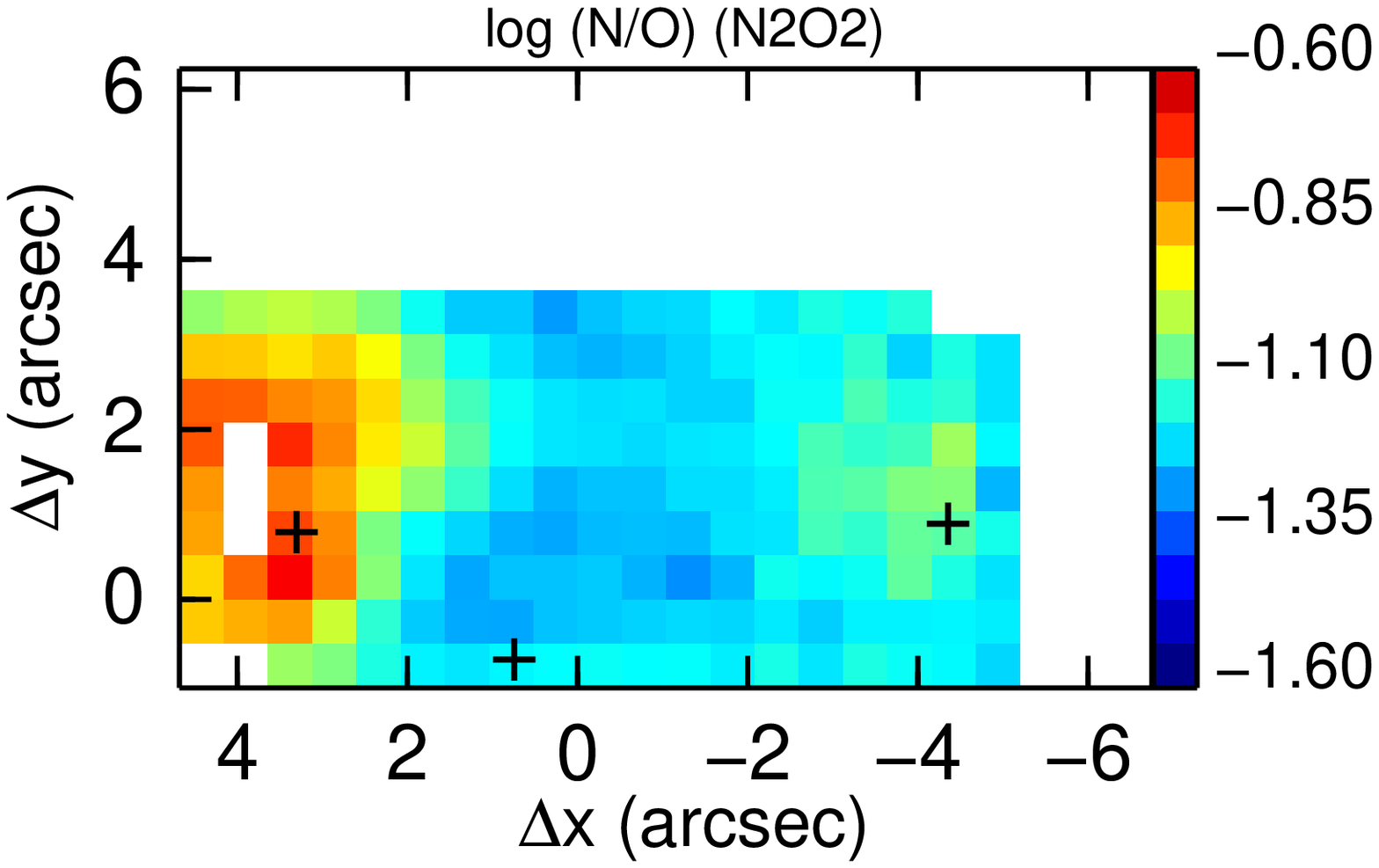}
\includegraphics[angle=0,width=0.48\textwidth,clip=, bb = 30 110 530 415]{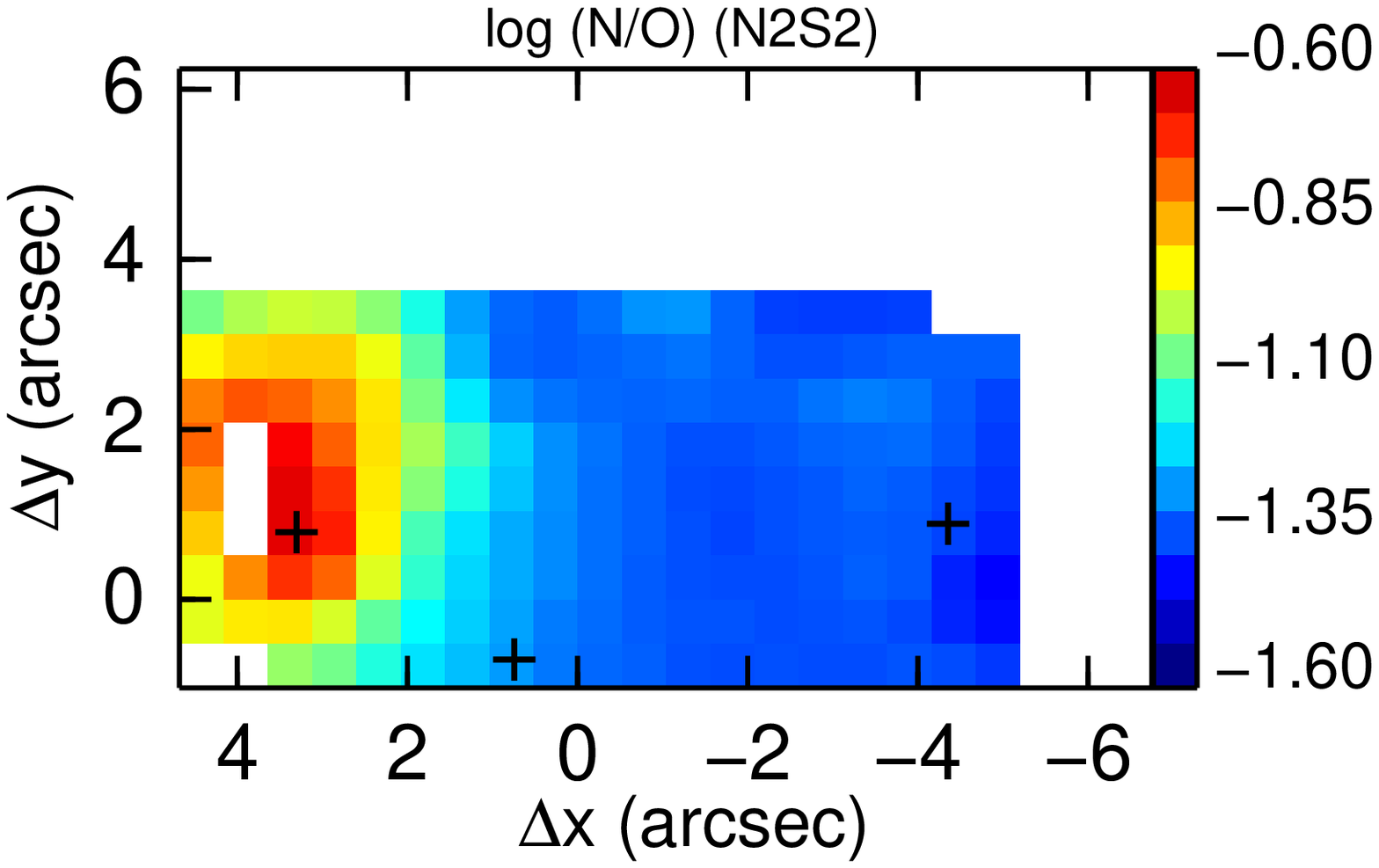}
   \caption[Predictions of strong-line based tracers for $N/O$. Relative abundances maps]{Predicted $N/O$ maps using the N2O2 (\emph{top}) and N2S2 (\emph{bottom}) tracers. Note that for an easier comparison with the predictions of the direct methods, maps are displayed using a scale covering the same range as the one presented in Fig. \ref{mapabun}.
 \label{compaabun}}
 \end{figure}

\subsection{Validity of strong line methods: $N/O$ relative abundance}

Ideally, metallicity and relative abundances for the ionized gas in galaxies should be derived in a direct manner. However, this requires the determination of $T_e$, which depends on the detection of faint lines like \oiii$\lambda$4363. In extragalactic astronomy, most of the time this is not feasible. Instead, certain combinations of strong emission lines with more or less well established empirical and/or theoretical calibrations should be used \citep[e.g.][]{per09,kew08}. A comparison between both methodologies on a spaxel-by-spaxel basis would be useful to test the reliability of the strong line-based methods and identify the cause of possible biases. For example, \citet{mon11} and \citet{rel10} found that metallicity tracers are modulated by the ionization structure. The best-known metallicity tracer would probably be R23 \citep{pag79}. Unfortunately, with a 12+$\log(O/H)\sim8.28$, \object{NGC~5253} falls at the turn-over of the metallicity-R23 relation and therefore, this tracer is not appropriate for metallicity determinations in this galaxy. Alternatively, two widely used tracers are $O3N2=\log(($\ohb$)/($\nha$))$ \citep{all79} and N2=\nha\ \citep{den02}. However, both involve \nii\ emission lines, and are therefore, affected by the variation of relative $N/O$ abundance across the galaxy. Thus, \object{NGC~5253} is not an adequate example to test the reliability of metallicity determinations based on strong optical lines.
Instead, since this galaxy presents a range of $N/O$ abundances, a different test for the corresponding strong line based tracers would be of high interest. This will be discussed in this section. As a baseline, we converted our line ratio maps into relative abundances using the expressions provided by \citet{per09}, that relate $N/O$ 
with N2O2 = $\log$(\nii$\lambda$6584/\oii$\lambda\lambda$3726,3729) and N2S2 =$\log$(\nii$\lambda$6584/\sii$\lambda\lambda$6717,6731) as defined by \citet{kew02b} and \citet{sab77}.

Maps for the estimated relative $N/O$ abundances are presented in Fig. \ref{compaabun}. They show that both tracers, N2O2 and N2S2, detect the main area with N-enhancement. Interestingly enough, only N2O2 is sensitive to the newly discovered area, associated to knot $\sharp$3. 

\begin{figure}[thb!]
   \centering
\includegraphics[angle=0,width=0.45\textwidth]{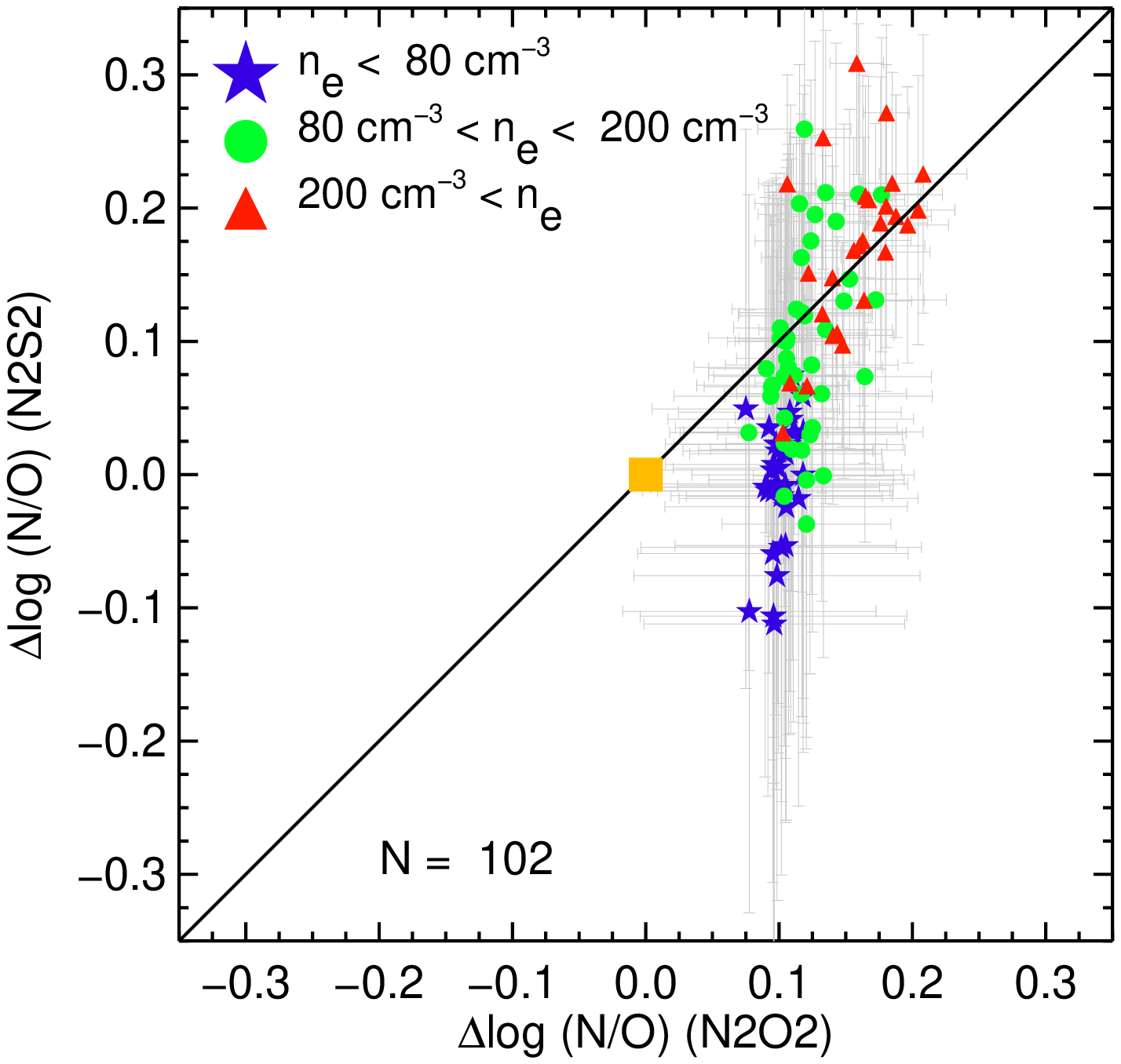}
  \caption[Metallicity and relative abundances maps]{Comparison between the residuals for the determination of $N/O$ based on the two strong line methods under consideration with respect to the direct one. The locus of equal residuals is indicated with a black line. An orange square marks the position where the three methods agree. Data were divided in three bins of $n_e$ as in Figs. \ref{compau} and \ref{compate}. 
 \label{comparesi}}
 \end{figure}

However, both tracers fail to predict the correct $N/O$ abundance. This is not unexpected since the utilized relations are  valid to interpret global tendencies in a statistically significant sample of galaxies, while $N/O$ abundances for an individual object can depart about $\pm0.3$ and $\pm0.5$~dex from this relation for the N2O2 and N2S2 tracers respectively \citep[see e.g. Figs. 10 and 11 in][]{per09}.

Figure \ref{comparesi} shows a comparison of the residuals between the strong line methods and the direct one.
Only those spaxels with an estimated uncertainty lower than 0.25~dex in both residuals were included in the comparison.
The good correlation between the residuals is consistent with both methods being affected by the same factors. Also, there is an increase of the residual with $n_e$. This also implies an increase with the other physical parameters since a comparison of the different maps presented throughout the paper shows that in general, larger $n_e$ corresponds to brighter areas and with larger $T_e$ and ionization strength. Given the spatial coincidence of these variations, disentangling the relative role of a given physical quantity as the cause of the variations in the residuals is not straightforward for this galaxy and will not be addressed here. 
Mean ($\pm$standard deviation) for the residuals are $\sim0.11 (\pm0.03)$ and $\sim0.02 (\pm0.12)$ for N2O2 and N2S2, respectively. Therefore, a method based on N2O2 is $\sim$4 times less sensitive to any variation of the physical/chemical properties than the one based on N2S2.

In summary, our comparison of the $N/O$ abundances derived using strong line methods with respect to direct measurements in the search for chemical inhomogeneities within a galaxy, supports the use of that based on N2O2 over that based on N2S2. The first method is sensitive to a wider range of $N/O$ abundances and is more stable against variations of physical conditions within the area of interest. 


\subsection{On the relationship between extra N and WR stars}

There are several works in the literature using long-slit observations that suggest a connection between nitrogen overabundance and the presence of WR stars based on the simultaneous detection of WR features and a higher than the expected $N/O$ in a given object \citep[e.g.][]{wal87,thu96,gus00,pus04,izo06b,hag06,per10,lop12}. Also, local associations between WR emission and N-enhancements  have been found by means of IFS based observations \citep[e.g.][]{jam09} supporting this connection. 
However, the general picture is far from this simple one-to-one association. For example, \citet{per11} studied a sample of BCD galaxies that are N-overabundant over large areas supporting the idea of enrichment caused by accretion of less chemically processed gas \citep[see also][]{keh08}. A similar result was found by \citet{amo12} for a set of BCD galaxies at higher redshift.
On top of that, resolution effects also play a important role. Already at distances of $\sim$25-40~Mpc the whole area covered here would have a projected size of the order of the resolution of typical ground based observations ($\sim$1\farcs0). 
%


We reported in \citetalias{mon10} areas with both extra-N and WR emission in \object{NGC~5253}, as for example, our knot $\sharp$1  \citep[widely discussed in the literature, e.g.][]{wal89,kob97,sch99,lop07} together with areas with WR emission but without any chemical anomaly (e.g. our knot $\sharp$2). Moreover, we report in Sec. \ref{secabun} a new area presenting an overabundance of nitrogen without any WR emission associated to it. 
This variety of options (i.e. WR emission and enhanced N, no WR emission and enhanced N, WR emission and no enhanced N, no WR emission and no enhanced N - i.e. the normal case) seems a more natural situation than a one-to-one association since the chemical enrichment of the warm ionized medium due to the material expelled by the massive stars is a complex process. In a simplified manner, this can be divided in three basic steps.

 Firstly, stars must bring their processed material to the surface. Recent stellar evolution models show how the inclusion of rotation favors the appearance of processed material at stages as early as the main sequence \cite[i.e. ][]{mae00,mae05,mae10}. However they are particularly powerful at the WR, or even Luminous Blue Variable, phases \citep[see][and references therein]{cro07}. Also, observational evidence of N-enhancement in WR ring nebulae has been reported \citep{fer12}.
 
 
  Secondly, this material should be expelled via stellar winds, first, and supernova explosions, afterwards. In the case of star clusters, where stars with a variety of initial masses have a variety of evolutionary paths, the yield of the different elements varies in a non-trivial manner with time \cite[e.g.][]{mol12}. An extra caveat arises for clusters with stellar masses of $\lsim10^4$~M$_\odot$ since in this regime the Initial Mass Function is not properly sampled and stochastic effects are important  \cite[e.g.][]{vil10}.
  
 Finally, this new material must be disseminated and then diluted in the warm ionized medium which, in due time, would reach a new chemical homogeneity. This is an even less trivial step since on top of the specific characteristics of a given star cluster, factors like the existence and characteristics  of neighbouring clusters, presence of cloudlets of gas, degree of the inhomogeneities, etc. are important to properly trace the evolution of the yielded material \citep[e.g.][]{ten96,dea02}.

Therefore, a unique evolutionary scenario to describe the path from the creation of the new nitrogen to its incorporation to the warm ionized medium seems unlikely. In the following we propose some  evolutionary paths for our  knots $\sharp$1-3. These scenarios do not intend to be more than reasonable suggestions that compile the constraints derived from our results and which could be tested with detailed modeling.

Regarding our knots $\sharp$2 and $\sharp$3, we hypothesize here about the possibility that  they constitute two snapshots of the same evolutionary path. According to \citet{har04} the main clusters associated to them are relatively similar in terms of mass and youth, although those in knot $\sharp$3 are slightly older\footnote{Knot $\sharp$2 corresponds to clusters 4 and 8 in \citet{har04}, with stellar masses of $\sim$2.7 and 1.3$\times10^4$~M$_\odot$ and ages of 1 and 5 Myr while knot $\sharp$3 corresponds to their cluster 3 and 5, with masses 4.2 and 2.1$\times10^4$~M$_\odot$ and ages of 8 and 11 Myr, respectively.}. Supporting this youth, candidates to supernova remnants have been found close to both knots \citep{lab06}. Moreover, only knot $\sharp$2 presented spectral features typical of Wolf-Rayet stars. In this evolutionary path, stars would expel their processed nitrogen at very early stages ($\lsim$5~Myr).
The difference between the age of the cluster in knot $\sharp$3 ($\sim$10~Myr) and the moment when the yielding of extra-N took place \citep[as soon as $\sim2.5$~Myr, according to the models of][]{mol12} sets an upper limit of $\sim$8~Myr for the duration of this process. If rotating O stars were playing a relevant role, this limit could be extended up to the age of clusters. A lower limit can be estimated under the most optimistic (and efficient) assumption: instantaneous incorporation of the yielded material to the warm ionized gas. If, in addition, we assume that the velocity of the ionized gas with respect to knot $\sharp$3 \citepalias[$\Delta$v$\sim$20~km~s$^{-1}$, see][]{mon10} traces the velocity under which the contamination of extra nitrogen propagates through the ISM, the area could be enriched in only $\sim$2~Myr. Therefore, the process of the cooling down and mixing of the yielded material with the ISM over an area of 40-50~pc in diameter should last between $\sim$2 and $\sim$8~Myr. Note that the moment when the newly created material has completely mixed with the previously existing gas, and any chemical inhomogeneity has been erased, should occur much later and cannot be delimited with these observations further than $\sim$10~Myr.

Knot $\sharp$1 should follow a different evolutionary path since at this location both WR features and nitrogen enrichment are found. The knot is associated to two very young and massive super star clusters \citep[e.g.][]{sch99,gor01,alo04}, embedded in a  very dense and compact nebula \citep{tur00}. In that sense, they can be seen as nascent clusters that have not managed yet to disperse the cloud of gas where they were born. This implies a very particular set of physical conditions for the warm interstellar medium.
Indeed, a giant molecular cloud associated to this region has been reported \citep{mei02} and we showed through this work that $n_e$ and $T_e$ associated to this region are relatively high, as well as the degree of ionization. Also, the region presents supersonic velocity widths and high extinction \citepalias{mon10}. Somehow, the combination of some of these particular conditions has made possible the incorporation of the newly created nitrogen at an earlier stage.

A third example mentioned in previous work provides a different evolutionary path.
\cite{gar12} gives an age of 3.5~Myr and a stellar mass of $\sim5\times10^3$~M$_\odot$ for their knot C in \object{NGC~6789}. Moreover, they report a relative abundance of nitrogen $\sim$0.2~dex larger than in their other apertures. No WR feature was detected for this knot (Garc\'{\i}a-Benito, private communication). Clearly, at this mass regime, the yield is dominated by stochastic effects. Given the lack of any WR detection one has to resort to rotating O stars as the cause of the N-enhancement. Contrary to what happened with our knot $\sharp$1 no particularly extreme physical conditions other than a complex inner dust structure were reported for its surrounding ISM. This leaves the open question of how this ISM managed to incorporate this new material in such a short time scale.

All in all, even if there seems to be a connection between WR emission and nitrogen enhancement, as supported by the fact that WR galaxies show an elevated $N/O$ relative to non-WR galaxies \citep{bri08}, local examples like \object{NGC~5253} and \object{NGC~6789}, where linear spatial resolutions of $\sim$20~pc arcsec$^{-1}$ can be achieved, show how this relationship is complex. Specifically, the case of the knots $\sharp$2 and $\sharp$3 in \object{NGC~5253}, where WR emission and N-enhancement are associated to different star clusters separated by only $\sim$90~pc, illustrates the possibility that in galaxies at distances $\gsim25$~Mpc, the spatial coincidence of WR emission and $N/O$ overabundances does not necessarily imply an intimate association  between them based on \emph{cause-effect} relationship. However, this does not reject the possibility that both of them were related due to a common external cause. As an example, in \object{NGC~5253} one could identify as this external cause the putative event that triggered the starburst (i.e. the past interaction with M~83).

\section{Conclusions}

We have carried out a detailed 2D study of the physical and chemical properties of the ionized gas in the central part of \object{NGC~5253}, a very nearby BCD. The area was mapped in a continuous manner with the ARGUS-IFU unit of FLAMES. This work represents the natural continuation of the one presented in \citetalias{mon10}. 
The different maps utilized along the paper as well as the reduced data cubes are available as FITS files from the authors.

The major conclusions can be summarized as follows:

1. Physical and chemical properties of the ionized gas associated to the main star clusters were derived by extracting spectra in apertures of 9-10~spaxels. Measurements associated to knots $\sharp$1 and $\sharp$2 agree in general with those previously reported. The existing discrepancies are associated to the assumed electron temperature.
Abundances for knot $\sharp$3 (not reported so far) are also provided. With the exception of the relative abundance for $N/O$ in knot $\sharp$1 and possible knot $\sharp$3, no chemical species seems overabundant in any of the selected apertures. 

2. Maps of the electron density based on four different tracers - namely \oii, \sii, \feiii, and \ariv\ line ratios  - were discussed. In all the cases, higher densities are associated to the main \ghiir. The joint analysis of these maps is consistent with a 3D stratified view of the nebula where the highest densities are located in the innermost layers while density decreases when going outwards.

3. The 2D $n_e$ structure for the two main kinematic components was also derived using the \oii\ lines as baseline. While the so-called broad component follows the picture described above, there is a change of structure for the narrow one. This fits well with the proposed scenario in \citetalias{mon10} where this component was associated to a shell of previously existing material that had been piled up by the outflow associated to the broad component. 

4. Maps for $T_e(\sii)$ and $T_e(\oiii)$ were derived. To our knowledge, this is the first time that a $T_e(\sii)$ map for an extragalactic object is presented. Also, we provided with one of the few examples of existing $T_e(\oiii)$ maps up to date. The joint interpretation of the information on the plane of the sky and along the line of sight is consistent with a $T_e$ structure in 3D with higher temperatures close to the main ionizing source surrounded by a colder and more diffuse component. This $T_e$ structure together with the lack of any strong broad component far from the main \ghiir\  \citepalias{mon10} is in accord with the lack of any clear evidence of shocks playing a dominant role.

5. Ionization structure was traced by means of the \oiiioii, \oii$\lambda\lambda$3726,3729/\hb, and \sii$\lambda\lambda$6717,6731/\ha\ ratios.
The two first of them predict similar ionization degree while the third one would be typical of lower ionization.
A possible 3D interpretation of both the observed structure in the maps for each individual ratio and the discrepancy of \sii$\lambda\lambda$6717,6731/\ha\ for individual spaxels is consistent with the lower ionization species (i.e. $S^+$) delineating the more extended diffuse component.

6. Maps for the 2D distribution of abundances for oxygen, neon, argon were derived. All of them are consistent with no chemical inhomogeneities.
The derived mean ($\pm$ standard deviation) oxygen abundance is $12+\log(O/H)=8.26\pm0.04$. The mean logarithmic relative abundances for argon and neon were $-0.65\pm0.03$ and $-2.33\pm0.06$, respectively.

7. In the same manner, a map for the 2D distribution of nitrogen was derived. \object{NGC~5253} typically presents a $\log(N/O)$ of $\sim-1.32\pm0.05$. However, there are two locations with enhanced $N/O$. The first one was
 already reported and characterized. With a $\log(N/O)\sim-0.95$, it occupies an elliptical area of about 80~pc$\times$35~pc and is associated to the two SSCs at the nucleus of the galaxy. The second one is reported here for the first time. It presents a $\log(N/O)\sim-1.17$ and it is associated to two moderately massive ($2-4\times10^4$~M$_\odot$) and relatively old ($\sim10$~Myr) clusters (knot $\sharp$ 3).
 
8. The map of $N/O$ relative abundance derived through the direct method was compared with those derived using strong line methods. The comparison supports a method based on N2O2 over a method based on N2S2
in the search of chemical inhomogeneities \emph{within} a galaxy since the first method is sensitive to a wider range of $N/O$ abundances and is more stable against variations of physical conditions within the area of interest.

9. We utilized the results on the localized detection of WR emission and nitrogen enhancement to compile and discuss the factors that affect the complex relationship between the presence of WR stars and $N/O$ excess. Even if there seems to be such a relationship, a unique scenario describing the path from the production of the new nitrogen to its incorporation into the warm ionized medium seems unlikely. In particular, we use the areas associated to knots $\sharp$2 and $\sharp$3 in \object{NGC~5253} as examples of WR emission and N-enhancement that would be perceived as spatially coincident at distances $\gsim25$~Mpc but that are not intimately associated in a cause-effect fashion. However, this does not reject the possibility that both of them are related due to a common external cause. 


\begin{acknowledgements}

This paper has benefited from fruitful conversations during the Workshop "Metals in 3D: New insights from Integral Field Spectroscopy". We would like to thank in particular to 
R. Garc\'{\i}a-Benito, B. James, M. Moll\'a, L. Smith, G. Tenorio-Tagle, and E. P\'erez-Montero, M. Rela\~no as well as to M. Westmoquette and R. Amor\'{\i}n with whom we shared stimulating discussions that helped to improve it.
We also thank the referee for the useful comments that have significantly improved the first submitted version of this paper.


Based on observations carried out at  the European Southern
Observatory, Paranal (Chile), programmes 078.B-0043(A) and
383.B-0043(A). This paper uses 
the plotting package \texttt{jmaplot}, developed by Jes\'us
Ma\'{\i}z-Apell\'aniz,
\texttt{http://dae45.iaa.csic.es:8080/$\sim$jmaiz/software}. This 
research made use of the NASA/IPAC Extragalactic 
Database (NED), which is operated by the Jet Propulsion Laboratory, California
Institute of Technology, under contract with the National Aeronautics and Space
Administration.
The STARLIGHT project is supported by the Brazilian agencies CNPq,
CAPES, FAPESP and by the France-Brazil CAPES-COFECUB programme.

A.~M.-I. is supported by the Spanish Research Council within the programme JAE-Doc, Junta para la Ampliaci\'on de Estudios, co-funded by the FSE.
This work has been partially funded by the projects 
AYA2010-21887 from the Spanish PNAYA,  CSD2006 - 00070  "1st Science with
GTC"  from the CONSOLIDER 2010 programme of the Spanish MICINN, and TIC114 Galaxias y Cosmolog\'{\i}a of the Junta de Andaluc\'{\i}a (Spain). 

\end{acknowledgements}


\bibliography{mybib_aa}{}
\bibliographystyle{./aa}

\end{document}